\documentstyle[12pt]{article}
\makeatletter

\@addtoreset{equation}{section}
\def\theequation{\arabic{section}.\arabic{equation}}

\def\section{\@startsection{section}{1}{\z@}{3.5ex plus 1ex minus
   .2ex}{2.3ex plus .2ex}{\large\bf}}
   \def\thesection{\arabic{section}}

\def\appendix{\setcounter{section}{0}
        \def\thesection{Appendix\ \Alph{section}}
        \def\theequation{\Alph{section}.\arabic{equation}}}

\newcommand{\beq}{\begin{eqnarray}}
\newcommand{\eeq}{\end{eqnarray}}
\newcommand{\eq}{eqnarray}
\newcommand{\bb}{\bibitem}

\newcommand{\be}{{\beta}}
\newcommand{\ci}{\cite}
\newcommand{\ga}{{\gamma}}
\newcommand{\Ga}{{\Gamma}}
\newcommand{\ep}{{\epsilon}}

\newcommand{\de}{{\delta}}
\newcommand{\De}{\Delta}

\newcommand{\om}{{\omega}}
\newcommand{\Om}{{\Omega}}
\newcommand{\pa}{{\partial}}
\newcommand{\no}{{\nonumber}}
\newcommand{\f}{\frac}
\newcommand{\ra}{\rightarrow}
\newcommand{\lra}{\leftrightarrow}
\newcommand{\eff}{\hbox{\scriptsize eff}}

\newcommand{\mic}{micro-canonical }
\newcommand{\can}{canonical }

\newcommand{\mlo}{\mbox{ln}}

\begin{document}

\topmargin 0pt

\oddsidemargin -3.5mm

\headheight 0pt

\topskip 0mm
\addtolength{\baselineskip}{0.20\baselineskip}
\begin{flushright}
\end{flushright}
\begin{flushright}
hep-th/0402173
\end{flushright}
\vspace{0.5cm}
\begin{center}
    {\large \bf  Testing
     Holographic Principle from Logarithmic and Higher Order Corrections to Black Hole Entropy}
\end{center}
\vspace{0.5cm}
\begin{center}
Mu-In Park\footnote{~Electronic address: muinpark@yahoo.com} \\
{ Department of Physics, POSTECH, Pohang 790-784, Korea}\\
\end{center}
\vspace{0.5cm}
\begin{center}
    {\bf ABSTRACT}
\end{center}
The holographic principle is tested by examining the
    logarithmic and higher order corrections to the
    Bekenstein-Hawking entropy of black holes. For the BTZ black
    hole,
    I find some disagreement in the principle
    for a holography screen at spatial infinity beyond the leading
    order, but a holography with the screen at the horizon does not,
    with an appropriate choice of a period parameter, which has been
    undetermined at the leading order, in Carlip's horizon-CFT approach
    for black hole entropy in any dimension. Its higher dimensional
    generalization is considered to see a universality of
    the parameter choice. The horizon holography from Carlip's is
    compared with several other realizations of a horizon
    holography, including induced Wess-Zumino-Witten model approaches
    and quantum geometry approach, but none of the these
    agrees with Carlip's, after clarifications of some
    confusions. Some challenging open questions are listed finally.

 \vspace{3cm}
\begin{flushleft}

1 Dec. 2004 \\
\end{flushleft}

\newpage

\section{Introduction}
In the recent ten years, there have been enormous studies on the
``holographic principle", which states that the world in the bulk
space-time is encoded on its boundary surface: holographic screen
\cite{Hoof:93}. In other words, this principle says, roughly
speaking,
\begin{\eq}
\label{holo}
 Z_X \sim Z_{\pa X}
\end{\eq}
with the partition functions $Z_X$ on the bulk space-time $X$ and
$Z_{\pa X}$ on its boundary $\pa X$. This has been now well-tested
when anti-de Sitter spaces are involved in the bulk, and some
appropriate conformal field theories are considered on the
infinite
boundary
in the context of string theory
(AdS/CFT) \cite{Ahar:99}. Similar correspondences for de-Sitter
\ci{Park:98,Bous:99,Stro:01} and flat spaces \ci{Bous:99,Son:02}
have been considered recently, but concrete realizations have not
been achieved yet.

On the other hand, when there are black holes in the bulk
space-time $X$, it seems that, according to 't Hooft and Susskind
\ci{Hoof:93}, collection of event horizons can be considered as
holographic screens also. Similarly, when there is the
cosmological horizon, this
can
then be a holographic screen also \ci{Sach:01}. There is a
strong evidence for this in the computation of the
Bekenstein-Hawking(BH) entropy from a conformal field theory
living on the {\it horizon}
\ci{Carl:95,Carl:97,Asht:98,Teit:96,Carl:99,Park:99,Park:02,Solo:99,Carl:02,Kang:04}.
Since the BH entropy is the exponent of the partition functions at
the ``leading order'', this result implies the relation
(\ref{holo}) holds at the leading order. However, this is in
contrast to Strominger's computation on the BTZ black hole entropy
from a conformal field theory at {\it spatial infinity}
\ci{Stro:98} that implies a holography (\ref{holo}) at spatial
infinity {\it at the same (leading) order}. So, at least for the
leading order, both the holography at spatial infinity and that of
the horizon give an identical result, and one can not distinguish
between these very different approaches.

In this paper, I show that, for the BTZ black hole, the two
holography prescriptions give different results, so these can be
differentiated beyond the leading order. I find that, remarkably,
there is some disagreement in
the holography at spatial infinity
beyond the leading order.
This may be
in contrast to
the
AdS/CFT
predictions.
On the other hand, the holography at the horizon may be satisfied
by choosing an appropriate period $P$, which has been arbitrary
for the leading-order computations, of the metric diffeomorphism
parameter $\xi^{\mu}$ in Carlip's horizon-CFT approach of black
hole entropy in any dimension \ci{Carl:99,Park:99,Park:02}. The
higher dimensional generalization is considered from the
Schwarzschild-anti-de Sitter black holes in $d \geq 4$ to see a
universality of the parameter choice. This horizon holography from
Carlip's is compared with several other realizations of a horizon
holography, including induced Wess-Zumino-Witten model approaches
for the Lorenzian and the Euclidean BTZ black holes; quantum
geometry approach
for the four-dimensional Schwarzschild and Schwarzschild-AdS black
holes. But, I find that none of these other approaches agrees with
Carlip's,
 after clarifications of some confusions. Some challenging open
questions are listed finally.

\section{General Logarithmic Corrections to the Bekenstein-Hawking
entropy}

I start by considering the density of states $\Omega(E)$ in the
{\it micro-canonical} ensemble, where the total energy $E$ is
fixed and always bigger enough than the $tolerance$ $\delta E$,
i.e., $\delta E \ll E$,
\begin{\eq}
\label{micro}
\Omega(E)&=&\sum _r \delta(E_r-E) \delta E  \\
\label{microZ}
 &=&\f{\delta E}{2
\pi} \int^{\infty}_{-\infty} d \be ' e^{\hat{\be}E} Z[\hat{\be}],
\end{\eq}
where $\hat{\beta}=\be +i \be'$, and $\be$ is an arbitrary
parameter which can be chosen at will. The partition function
$Z[\hat{\be}]$ in the {\it canonical} ensemble for the complex
$\hat{\be}$ is defined as
\begin{\eq}
\label{Z:canonical}
 Z[\hat{\be}]=\sum_r e^{-\hat{\be} E_r}.
\end{\eq}
On the other hand, from the Laplace transformation for $\Om(E)$ of
(\ref{micro}), this can be alternatively expressed as
\begin{\eq}
\label{Laplace}
 Z[\hat{\be}]=\int^{\infty}_{-\infty} \f{dE}{\delta E}
\Omega(E)e^{-\hat{\be}E}.
\end{\eq}
However, note that the existence of the canonical partition
function depends on the convergence of the Laplace transform
(\ref{Laplace}): When the Laplace transform does not converge as
occurs in the Schwarzschild black hole, where the heat capacity is
negative such as it is thermodynamically unstable, one must return
to the original definition (\ref{micro}) to compute $\Om (E)$
\ci{Hawk:76,Brow:94}; but, for a thermodynamically stable system,
one can always use the formula (\ref{microZ}). Moreover, in this
computation it is important to note that {\it only the $\beta'
\leftrightarrow -\be'$ symmetric terms in the integrand
$e^{\hat{\be} E} Z[\hat{\be}]$ of (\ref{microZ}) are relevant by
construction}; this property has a crucial role, as can be seen in
the later part, when one computes (\ref{microZ}) perturbatively
beyond the first correction.

In the stationary phase method, the integral (\ref{microZ}) can be
perturbatively evaluated around $\be'=0$, where the usual energy
formula at the equilibrium temperature $T=\be^{-1}$
\begin{\eq}
E=-\left.\f{\pa \mbox{ln} Z[\hat{\be}]}{\pa {\hat
\be}}\right|_{\be'=0}
\end{\eq}
is satisfied, as follows
\begin{\eq}
\label{Omega} \Omega (E)= e^{{\be}E} Z[{\be}] \times \f{\delta
E}{2 \pi} \int^{\infty}_{-\infty} d \be ' \mbox{exp} \left\{
-\f{1}{2} B_2 \be'^2 +\sum^{\infty}_{n\geq 3} \f{1}{n !} B_n (i
\be')^n \right\} ,
\end{\eq}
where
\begin{\eq}
B_n=\left.\f{\pa^n \mbox{ln} Z[\hat{\be}]}{\pa {\hat
\be}^n}\right|_{\be '=0}.
\end{\eq}
It is a standard result \ci{Reif:86} that the (micro-canonical)
entropy $S=\mbox{ln} \Om (E)$ is computed as
\begin{\eq}
\label{S} S=S_c-\f{1}{2} \mbox{ln} \left[ \f{2 \pi C_x T^2}{(\de
E)^2 }\right] +\mbox{(higher order~terms)},
\end{\eq}
where
the second term comes from the Gaussian integral
$\int^{\infty}_{-\infty} d \be ' \mbox{exp} ( -B_2 \be'^2/2)=
\sqrt{2 \pi/B_2}$ and
$B_2=C_x T^2$ ($C_x$ is the specific heat with a fixed extensive
parameter $x$). And
\begin{\eq}
\label{Sc}
 S_c=\be E +\mbox{ln} Z[\be]
\end{\eq}
is the entropy defined in the canonical ensemble; I shall mean
`entropy' as the micro-canonical entropy later on unless it is
stated otherwise.
The `higher order terms' represents the correction terms from the
integral $\int^{\infty}_{-\infty} d \be ' \mbox{exp} [
\sum^{\infty}_{n\geq 3}  B_n (i \be')^n/ {n !}]$.
In the usual statistical mechanics, $T$ is independent on the
system size, i.e., intensive variable, such as $T^2$ term in the
logarithmic term of (\ref{S}) can be neglected in the
thermodynamic limit since $C_x$ is an extensive quantity, which
can be arbitrarily large like as $E$ and $S$ \footnote{ I thank K.
Huang for a helpful comment on this.}. But still, the term of
tolerance $\de E$ is neglected since $\de E$ is defined to be
smaller enough than any available energy fluctuation, i.e., $\de E
\ll \De E=C_x T^2$ for the mean-square fluctuation in the energy
\ci{Huan:63}.

However, in the application of this statistical mechanical
formulation into black hole thermodynamics by the Euclidean
path-integral approach \ci{Gibb:77,Brow:91} $T$ depends on the
system size, i.e., the horizon size, such as the $T^2$ term can
not be neglected anymore in the logarithmic term of (\ref{S}).
This is one of the big differences between the usual
thermodynamics and black hole thermodynamics. But one can still
assume safely {\it $\de E  \ll$ any available energy fluctuation},
by considering {\it large} black holes. Moreover, the relation
(\ref{S}) between the \mic entropy and the \can entropy $S_c$
shows a perturbative correction \ci{Das:02} to the BH entropy
\footnote{If I include quantum corrections, I need to generalize
the Einstein gravity to a higher curvature gravity with a
$possible$ UV-finite $ln A$ correction. But this would not affect
our result for large black holes. See discussion 1 in section 8
for details. } \ci{Gibb:77,Brow:91}
\begin{\eq}
\label{SBH}
 S_c\equiv S_{BH}=\f{A}{4 \hbar G},
\end{\eq}
where $A$ is the horizon area, and $G$ is the Newton's constant;
when there is rotational degrees of freedom, one can generalize
straightforwardly to the grand-canonical partition function and
its associated density of states, but this will not be considered
in this paper.

\section{General Logarithmic Corrections to the BH entropy from
the Cardy Formula}

There is an analogous situation in the two-dimensional conformal
field theory on a two-torus, where one can perturbatively evaluate
the density of states. To see this, I begin with the partition
function of the conformal field theory on a torus
\ci{Card:86,Carl:98}\footnote{Compare to the derivation of Carlip
\ci{Carl:98}, who used a slightly different partition function
$\hat{Z}[\tau,\bar{\tau}]=Tr e^{2 \pi i \tau L_0} e^{-2 \pi i
\bar{\tau} \bar{L}_0}$, which is {\it not} modular invariant. But,
the result for the density of states $\rho(\De)$ is the same.}
\begin{\eq}
\label{CFTZ}
 Z[\tau,\bar{\tau}]=Tr e^{2 \pi i \tau (L_0-\f{c}{24})}
e^{-2 \pi i \bar{\tau} (\bar{L}_0-\f{\bar{c}}{24})}
\end{\eq}
with the modular parameters $\tau,\bar{\tau}$ and the Virasoro
generators $L_m,\bar{L}_m$ on the ``plane'' with central charges
$c, \bar{c}$, with the algebras in the standard form,
\begin{\eq}
\label{Virasoro}
&&[L_m , L_n]=(m-n)L_{m+n} +\f{c}{12} m (m^2-1) \de_{m+n,0}, \no \\
&&[\bar{L}_m, \bar{L}_n]=(m-n) \bar{L}_{m+n} +\f{\bar{c}}{12} m (m^2-1) \de_{m+n,0}, \no \\
&&[L_m, {\bar L}_n]=0.
\end{\eq}
The density of states $\rho(\De, \bar{\De})$ for the eigenvalues
$L_0=\De, \bar{L}_0=\bar{\De}$ is given as a contour integral (I
suppress the $\bar{\tau}$-dependence for simplicity, but the
computation is similar to the $\tau$-part)
\begin{\eq}
\label{rho}
 \rho(\De) =\int_C d\tau~ e^{-2\pi i (\De -\f{c}{24})
\tau } Z[\tau],
\end{\eq}
where the contour $C$ encircles the origin in the complex $q=e^{2
\pi i \tau}$ plane \ci{Carl:97}, and the tolerance is given by
$\de \De=1$ by definition. The general evaluation of this integral
would be impossible unless $Z[\tau]$ is known completely. But, due
to the modular invariance of (\ref{CFTZ}), one can easily compute
its asymptotic formula through the steepest descent approximation.
In particular, (\ref{CFTZ}) is invariant under $\tau \ra -1/\tau$
\ci{Card:86} such that
\begin{\eq}
Z[\tau] =Z[-1/\tau]=e^{-2 \pi i (\De_{\hbox{\scriptsize
min}}-\f{c}{24}) \tau }\tilde{Z}[-1/\tau],
\end{\eq}
where $\tilde{Z}[-1/\tau]=Tr e^{-2 \pi i
(L_0-\De_{\hbox{\scriptsize min}})/ \tau}$ approaches a constant
value $\rho(\De_{\hbox{\scriptsize min}})$ as $\tau \ra i 0_+$,
which defines the steepest descent path for a ``real'' value of
$\De \geq \De_{\hbox{\scriptsize min}}$. With the help of this
property, (\ref{rho}) is evaluated as, by expanding the integrand
around the steepest descent path $\tau_*$,
\begin{\eq}
\label{rho-ser}
 \rho(\De) &=&\int_C d\tau~ e^{\eta(\tau)} \tilde{Z}[1/\tau] \\
&=&e^{\eta(\tau_*)}  \tilde{Z}[-1/\tau^*] \times \int_C d \tau
~\mbox{exp} \left\{ \f{1}{2} \eta^{(2)} (\tau-\tau_*)^2
+\sum^{\infty}_{n\geq
3} \f{1}{n !} \eta^{(n)} (\tau-\tau_* )^n \right\} \no \\
 &&\times \left[ 1+\sum^{\infty}_{m \geq 1} \f{1}{m !} \tilde{Z}^{-1}
\tilde{Z}^{(m)} (\tau-\tau_*)^m \right], \label{rho-expand}
\end{\eq}
where $\eta(\tau)=-2 \pi i \De_{\eff} \tau +{2 \pi i}
c_{\eff}/({24}{\tau})$, which dominates $\tilde{Z}[1/\tau]$ in the
region of interest, gets the maximum
\begin{\eq}
\eta(\tau_*) &=&2 \pi \sqrt{\f{c_{\eff} \De_{\eff}}{6}}
\end{\eq}
with
\begin{\eq}
\tau_*&=&i \sqrt{\f{c_{\eff}}{24 \De_{\eff}}}
\end{\eq}
when
\begin{\eq}
\label{cond} \De_{\eff} \gg \f{c_{\eff}}{24}
\end{\eq}
is satisfied. Here, $\eta^{(n)}= ({d^n \eta}/{d
\tau^n})|_{\tau=\tau_*}, ~\tilde{Z}^{(m)}=({d^n \tilde{Z}}/{d
\tau^n})|_{\tau=\tau_*} $, and $c_{\eff}=c-24
\De_{\hbox{\scriptsize min}},~ \De_{\eff}=\De-{c}/{24}$;
$\De_{\hbox{\scriptsize min}}$ is the minimum of $\De$. Here, I am
assuming ``$c_{\eff}, \De_{\eff}>0$'' since, otherwise, the saddle
point approximation is not valid for $real$ valued
$c_{\eff},~\De_{\eff}$. Then, in the limit of $\ep \ra \infty$
with $\tau_*=i/\ep$, the higher order correction terms in the
bracket of (\ref{rho-expand}) are exponentially suppressed as
$e^{- 2 \pi \ep (\De-\De_{\hbox{\scriptsize min}})}$, hence
(\ref{rho-expand}) is simplified as, up to the exponentially
suppressing terms,
\begin{\eq}
\label{rho-ser'}
 \rho(\De) =e^{2 \pi \sqrt{c_{\eff} \De_{\eff}/6}}
  \times \int_C d\tau~ \mbox{exp}
\left\{ \f{1}{2} \eta^{(2)} (\tau-\tau_*)^2 +\sum^{\infty}_{n\geq
3} \f{1}{n !} \eta^{(n)} (\tau-\tau_* )^n \right\},
\end{\eq}
where I have used $\tilde{Z} [i \infty] =1$. This is known as the
Cardy formula \ci{Card:86}. Note that here I need $c_{\eff}
\De_{\eff} \gg 1$ in order that the approximation is reliable,
i.e., $e^{\eta(\tau_*)}$ dominates in the integral of
(\ref{rho-ser}), as well as the condition (\ref{cond}) such as
$\tilde{Z}[-1/\tau]$ is slowly varying near $\tau_*$. Then, the
entropy $S_{CFT}=\mbox{ln} ~\rho(\De)$ of this $\tau$ sector
becomes
\ci{Carl:98}
\begin{\eq}
\label{SCFT}
 S_{CFT}=S_0+\f{1}{4} \mbox{ln}\left[ \f{c_{\eff}}{96
\De_{\eff}^3}\right] +(\mbox{h.o.t.}),
\end{\eq}
where
\begin{\eq}
\label{S0}
 S_0 =2 \pi \sqrt{c_{\eff} \De_{\eff}/6},
\end{\eq}
the second term comes from the contour integral $\int_C d\tau~
\mbox{exp} (  \eta^{(2)} (\tau-\tau_*)^2/2)=\sqrt{2
\pi/|\eta^{(2)}|}$,
and `(h.o.t)' denotes the higher order terms from $\int_C d\tau~
\mbox{exp} [\sum^{\infty}_{n\geq 3}  \eta^{(n)} (\tau-\tau_* )^n
/{n !}]$. Now, by recovering $\bar{\tau}$-part also, the total
entropy becomes
\begin{\eq}
\label{SCFT2} S_{CFT^2}=S_0+\bar{S}_0+\f{1}{4} \mbox{ln} \left[
\f{c_{\eff}\bar{c}_{\eff}}{(96)^2 \De_{\eff}^3
\bar{\De}_{\eff}^3}\right] +(\mbox{h.o.t.}).
\end{\eq}
This shows a similar logarithmic correction term as the
first-order correction in the bulk (\ref{S}). But, it seems in a
priori that these CFT entropies $S_{CFT, CFT^2}$ need not be
related to the bulk entropy $S$, which has been computed in a
completely different context. However, in the context of the
holographic principle
this ``might'' be related if one considers the conformal field
theory in the above computation lives on a holographic screen.
This remarkable connection does exit at the leading order, as has
been first shown by Strominger \ci{Stro:98}; he has shown that the
CFT entropy associated with the asymptotic isometry $SO(2,2)$ at
spatial infinity of the BTZ black hole agrees with the BH entropy,
which is the entropy in the bulk at the leading order. This has
been considered as a concrete example of the AdS$_3$/CFT$_2$.
Moreover, a similar coincidence has been observed in the
dS$_3$/CFT$_2$ context also, where there is the cosmological
horizon instead, though its associated Cardy formula has not been
proved yet \ci{Park:98}\footnote{This has been proved only
recently by the author \ci{Park:04}. I will discuss on this
later.}, and though it has a conceptual problem of the meaning of
the holographic screen beyond the casually connected region.

On the other hand, recently it has been shown that there is also a
conformal filed theory on the event horizon for arbitrary black
holes in any dimension, and its associated CFT entropy agrees with
the BH entropy at the leading order
\ci{Carl:99,Solo:99,Carl:02,Kang:04}. This latter approach is
conceptually better than the former one of Strominger for the
following reasons. First, in the former case, as noted by Carlip
\ci{Carl:99,Carl:02}, one can $not$ distinguish the BTZ black hole
with a point ``star'' with some appropriate mass and spin sitting
at the origin of AdS space \ci{Dese:84}; this is in contrast to
the holographic principle, where the data on a screen should
distinguish the black holes from the point star by definition;
however, this is automatically satisfied for the latter approach
by construction. Second, in de-Sitter case, there is an additional
conceptual problem that the screen is located in the
casually-disconnected region as I mentioned above. However, this
problem does not appear in the latter approach either. But,
unfortunately one can not distinguish those two very different
holography schemes at the leading-order computations. This raises
the following
question of the higher order corrections:
{\it Can we determine the correct one by assuming the holographic
principle $Z_{X} \sim Z_{\pa X}$ even at the higher orders ?}

\section{Test I: The BTZ Black Hole}

\subsection{In the Bulk:}

The BTZ black hole solution \ci{Bana:92} is the solution to the
vacuum Einstein equation in 2+1 dimensions with a negative
cosmological constant $\Lambda=-1/l^2$. The metric is given by
\begin{\eq}
ds^2=-N^2 dt^2 +N^{-2} dr^2 + r^2 (N^{\phi} dt +d \phi)^2,
\end{\eq}
where
\begin{\eq}
N^2=-8G M+\f{r^2}{l^2}+\f{16 G^2 J^2}{r^2},~ N^{\phi}=-\f{4 G
J}{r^2}.
\end{\eq}
$M$ and $J$ are the ADM mass and angular momentum of the black
hole ($M>0,|J|\leq ML)$, and there are inner and outer horizons at
\begin{\eq}
r_{\pm}=2 \sqrt{G} l \sqrt{M\pm \sqrt{M^2-(J/l)^2}}.
\end{\eq}
The BH entropy, the Hawking temperature, and the angular velocity
of the horizon are given by, respectively,
\begin{\eq}
S_{BH}&=&\f{2 \pi r_+}{4 G}, \\
T&=&\f{\kappa}{ 2\pi}= \f{r_{+}^2-r_{-}^2}{2 \pi l^2 r_{+}},\\
\Omega&=&\f{4 G J}{ r_+^2}.
\end{\eq}
The heat capacity $C_J=(\pa M/\pa T)_J$ for `$J$=fixed' becomes,
after some computation,
\begin{\eq}
C_J=\f{4 \pi r_+ (r_+^2-r_-^2)}{8 G (r_+^2 +3r_-^2)}.
\end{\eq}
Note that $C_J$ is always positive since $r_+>r_-$ for
non-extremal black holes ($T>0$) such as the canonical ($T=$fixed)
or grand-canonical ($T=\Omega=$fixed) ensemble exists without
considering the equilibrium of black holes with a {\it
hypothetical~enveloping-box}, in contrast to the black hole
solutions in asymptotically flat space \ci{Hawk:83}. But, since I
am interested in the $large$ black holes of $r_+ \gg l$, as
appropriate in our perturbative computation, the canonical
ensemble, where $\Om$ is negligible, is enough. Then from
\begin{\eq}
C_J T^2 &=&\f{(r_+^2-r_-^2)^3}{8 \pi G l^4 r_+ (r_+^2 +3 r_-^2)}
\no
\\
&\approx & \f{(8 G)^2}{4^3 \pi^4 l^4} S^3_{BH}
\end{\eq}
in the $r_+ \gg l$ regime, one can easily find the
entropy in the bulk as \footnote{ The ``ln$S_{BH}$'' corrections
were known earlier in the one-loop corrections due to quantum
fields on the black hole background in
Refs.\ci{Furs:95,Frol:96,Mann:98} (I thank R. B. Mann for
informing this); on the contrary, our corrections are essentially
due to the fluctuations of the black hole metric itself even
without the matters. However, the coeffients are not universal,
but depend on the matter contents. Furthermore, its connection to
the CFT's logarithmic correction was also studied \ci{Solo:98} in
the context of string/black-hole correspondence \ci{Suss:93},
where the string's ends are frozen on the horizon.}, from
(\ref{S}),
\begin{\eq}
\label{Sbulk:BTZ} S&=&S_{BH}-\f{1}{2} \mbox{ln}\left[ \f{S^3_{BH}
(8 G)^2}{32 \pi^3 l^4
(\de E)^2 }\right]+(\mbox{h.o.t.}) \no\\
&=&S_{BH}-\f{3}{2} \mbox{ln} S_{BH} +\f{1}{2} \mbox{ln}[32 \pi^3
l^4 (\de E)^2 /(8 G)^2] +(\mbox{h.o.t.}).
\end{\eq}

\subsection{At Spatial Infinity:}

Now, let us compare with the CFT computation from Cardy's formula
on a boundary. In this subsection first, let us consider the
screen at spatial infinity
as in
the usual AdS/CFT. To this end, I note that the associated
Virasoro algebra has two independent sectors as in
(\ref{Virasoro}), which have eigenvalues of $L_0,\bar{L}_0$ and
central charges \ci{Brow:86,Stro:98,Oh:98,Park:98,Park:02} as
\begin{\eq}
\label{De:BTZ}
\De_{\eff},\bar{\De}_{\eff} &=&\f{(r_+ \pm r_-)^2}{ 16 G l} \no \\
&\approx& \f{8 G}{32 \pi^2 l} ~S_{BH}^2, \\
\label{cBTZ}
 c=\bar{c}&=&\f{12 l}{8 G},
\end{\eq}
where $r_+\gg l$ is considered in (\ref{De:BTZ}). Then, one can
easily find the
entropy for the CFT, from (\ref{SCFT2}), as
\begin{\eq}
S_{CFT^2}=S_{BH}-3~ \mbox{ln} S_{BH} +\mbox{ln}[64 \pi^3 l^2 /(8
G)^{2}] +(\mbox{h.o.t.}),
\end{\eq}
where I have used
\begin{\eq}
S_0+\bar{S}_0&=&2 \pi \sqrt{c_{\eff} \De_{\eff}/6}+2 \pi
\sqrt{\bar{c}_{\eff} \bar{\De}_{\eff}/6} \no \\
&=&S_{BH}
\end{\eq}
with a choice $\De_{\min}=0$ \ci{Stro:98} \footnote{ This vacuum
is crucial in obtaining the correct BH entropy even when a scalar
field is coupled. See Ref.\ci{Park:0403}. }. This clearly shows
the factor of 2 mismatch,
with respects to the bulk result (\ref{Sbulk:BTZ}), at the
logarithmic order for a large black hole, i.e., large
$S_{BH}$.\footnote{A similar mismatch has been noted, for the
first time, by Carlip \ci{Carl:00} in a different context of
quantum geometry. But, as will be explained in later sections, his
observation was due to some misunderstanding of the entropy in
quantum geometry \ci{Kaul:00}. Moreover, he has not considered its
implication to the
holographic principle
since  the quantum geometry
approach also concerns about the ``horizon'' states, and the
associated ``bulk'' entropy is not considered.
} Hence, one finds that: \\

{\it There is some disagreement in the
holographic
correspondence
between
 3 dimensional
 gravity
 and 2 dimensional CFT, associated with
 the asymptotic isometry group
$SO(2,2)$ at the spatial infinity,
beyond the leading order}.\\

Although there is no higher dimensional analogy \ci{Henn:85}, this
higher order
disagreement of the
holographic principle in three dimensions
might
be far-reaching consequence because of its frequent
appearance in the many higher dimensional black holes in string
theory \ci{Hyun:97} as a sub-sector.

\subsection{At the Horizon:}

At the horizon $r_+$, the associated CFT has several different
features compared to that for spatial infinity. First of all,
there is only ``one'' copy of the Virasoro algebra instead of the
two copies (\ref{Virasoro}) at spatial infinity, though its origin
is at the ``classical'' level like as in spatial infinity
\ci{Carl:99}; the intuitive understanding of this is not clear to
us, but this has been observed in various different contexts
\ci{Solo:99,Carl:02,Kang:04} also. Second, the Virasoro algebra at
the horizon is quite universal for arbitrary black holes and any
dimension (except the extremal black holes and two-dimensional
black holes), in contrast to the Virasoro algebra for $AdS_3$
space at spatial infinity
\footnote{There seems to be a very strong hint on the universality
of $AdS_3$ Virasoro algebra for $AdS_d$ also.
For a recent achievement in the special context of string theory
and pp-waves see Ref.\ci{Bana:00}.  But there has been no general
proof for more general contexts yet.}.
 Though the ``most'' general from of the proper
boundary conditions near the horizon is not clear, for a ``quite''
general form of the boundary conditions near the horizon and an
appropriate choice of $\De_{\hbox{\scriptsize min}}$
($\De_{\hbox{\scriptsize min}}=0$ in our case), one obtains a
Virasoro algebra \ci{Carl:99,Carl:00} \footnote{There are some
mismatches in $c_{\eff}$ and $\De_{\eff}$ with other alternative
approaches \ci{Carl:02,Kang:04}. But, presumably similar
logarithmic corrections would be obtained also if the vacuum is
chosen properly \ci{Jing:00}.}
\begin{\eq}
\label{horizonCFT}
c_{\eff}&=&\f{3 A}{2 \pi G} \f{ 2 \pi}{ \kappa P}, \no \\
\De_{\eff}&=&\f{ A}{16 \pi G} \f{ \kappa P}{2 \pi},
\end{\eq}
where $A$ is the horizon area, $\kappa$ is the surface gravity,
and $P$ is the periodicity of the diffeomorphism parameter
$\xi^t$. Here, note that ``$\kappa P/(2 \pi)$'' terms of
(\ref{horizonCFT}) cancel in the computation of the leading term
of the CFT entropy (\ref{S0})
\begin{\eq}
S_0=2 \pi \sqrt{c_{\eff} \De_{\eff}/6} =\f{A}{4 G},
\end{\eq}
which is the BH entropy, such as $P$ is $not$ determined at the
leading order.

However, an explicit $P$-dependent term appears for higher order
corrections as follows, from (\ref{SCFT}),
\begin{\eq}
\label{SCFT:ln}
 S_{CFT}=S_{BH}-\f{1}{2} \mbox{ln} S_{BH}
-\mbox{ln} (\kappa P)+\f{1}{2} \mbox{ln} (8 \pi^3)+
(\mbox{h.o.t.}).
\end{\eq}
Now, by using
\begin{\eq}
\kappa &=&\f{r_+^2-r_-^2}{l^2 r_+} \no \\
&\approx &\f{8 G}{4 \pi l^2} S_{BH}
\end{\eq}
for the large black holes of $r_+ \gg l$ (with a finite $J$), one
finally obtains
\begin{\eq}
S_{CFT}=S_{BH}-\f{3}{2} \mbox{ln} S_{BH} -\mbox{ln} P+\f{1}{2}
\mbox{ln} (96 \pi^5 l^4/(8G)^2 )+ (\mbox{h.o.t.}).
\end{\eq}
Hence, one finds that the bulk entropy $S$ of (\ref{Sbulk:BTZ})
matches with its boundary entropy at the horizon $S_{CFT}$ if one
choose $P$ as
\begin{\eq}
P\approx 8 \pi/\de E,
\end{\eq}
which is a universal constant, independent of $A$. It is
interesting to note that $c_{\eff}$ becomes a universal constant
also
\begin{\eq}
\label{clog}
 c_{\eff}=c \approx \f{3 l^2 \de E}{4G},
\end{\eq}
while $\De_{\eff}\approx ({2G }/{\de E l^2 \pi^2}) S_{BH}^2$, with
a choice $\De_{\hbox{\scriptsize min}}=0$.

\section{Test  II : Schwarzschild-Anti-de Sitter black holes (d
$\geq 4$)}

As the higher dimensional generalization of the analysis of the
previous section, let us consider a higher dimensional
Schwarzschild black hole solution with the (negative) cosmological
constant $\Lambda=-(d-1)(d-2)/2l^2$ (Schwarzschild-AdS). The
metric is given by
\begin{\eq}
ds^2=-N^2 dt^2 +N^{-2} dr^2 + r^2 d \Om_{d-2}^2,
\end{\eq}
where
\begin{\eq}
N^2=1-\f{16\pi G M}{(d-2) \Om_{d-2}r^{d-3}}+\f{r^2}{l^2}.
\end{\eq}
$M$ is the mass of the black holes ($M>0)$, and $d \Om^2_{d-2}$ is
the line element on the unit sphere $S^{d-2}$
\ci{Hawk:83,Myer:86,Asht:84}, and the horizon radius increases as
the mass M as follows
\begin{\eq}
\label{M:adS-Sch}
 \f{16 \pi {G} {M}}{(d-2)
\Om_{d-2}}=r_{+}^{d-3} \left(1+\f{r_+^2}{l^2}\right),
\end{\eq}
where $\Om_{d-2}$ is the area of the unit $S^{d-2}$: $\Om_{d-2}=2
\pi^{(d-1)/2}/\Ga((d-1)/2)$. The BH entropy, the Hawking
temperature, and the heat capacity of the horizon are given by,
respectively,
\begin{\eq}
S_{BH}&=&\f{\Om_{d-2} r_+^{d-2}}{4 G}, \\
T&=&\f{1}{4 \pi r_+} \left(\f{d-1}{l^2} r_{+}^2 +(d-3)\right),\\
C_J&=&\f{(d-2)\Om_{d-2} r_+^{d-2}}{4 G }
\left(\f{(d-1)r_+^2/l^2+(d-3)}{ (d-1)r_+^2/l^2 -(d-3)}\right).
\end{\eq}
Note that $C_J$ is $not$ always positive, but becomes negative for
the small black hole of $r_+ <\sqrt{(d-3)/(d-1)}l$. However, this
does not matter in our case since I am only interested in the
large black hole of $r_+ \gg l$  such that our perturbative
expansion of black hole entropy is meaningful. Then from
\begin{\eq} C_J T^2 &=& \f{(d-2)\Om_{d-2} r_+^{d-4}}{4 G (4
\pi)^2 } \f{[(d-1)r_+^2/l^2+(d-3)]^3}{ [(d-1)r_+^2/l^2 -(d-3)]}\no
\\
&\approx & \f{(d-2)(d-1)^2}{(4\pi)^2 l^4}\left(\f{4 G}{\Om_{d-2}}
\right)^{2/(d-2)} S_{BH}^{d/(d-2)}
\end{\eq}
for the large black hole, one can easily find the
entropy in the bulk \ci{Das:02}, from (\ref{S}),
\begin{\eq}
\label{Sbulk:AdS-Sch} S&=&S_{BH}-\f{d}{2(d-2)} \mbox{ln}
S_{BH}+\f{1}{2} \mbox{ln}\left[\f{8 \pi l^4 (\de E)^2
}{(d-2)(d-1)^2}\left(\f{\Om_{d-2}}{4 G}\right)^{2/(d-2)} \right]
+(\mbox{h.o.t.}).
\end{\eq}

Now, since there is no CFT analogue at spatial infinity, as far as
I know, let us consider only the CFT at the horizon. There, the
formulas of the eigenvalues $L_0,\bar{L}_0$ and the central
charges (\ref{horizonCFT}) are valid for arbitrary higher
dimensions, such as one has the same ``formal'' higher order
corrections as (\ref{SCFT:ln}) also. But now, by using
\begin{\eq}
\kappa=2 \pi T&=&\f{1}{2 r_+} \left(\f{d-1}{l^2} r_{+}^2
+(d-3)\right) \no \\
&\approx& \f{(d-1)}{2 l^2}\left(\f{4 G}{\Om_{d-2}}
\right)^{1/(d-2)} S_{BH}^{1/(d-2)}
\end{\eq}
for the large black hole, one can finally obtains
\begin{\eq}
\label{SCFT:AdS-Sch} S&=&S_{BH}-\f{d}{2(d-2)} \mbox{ln}
S_{BH}-\mbox{ln} P+\f{1}{2} \mbox{ln}\left[\f{32 \pi^3 l^4 (\de
E)^2 }{(d-1)^2}\left(\f{\Om_{d-2}}{4 G}\right)^{2/(d-2)} \right]
+(\mbox{h.o.t.}).
\end{\eq}
Hence, one finds that the bulk entropy $S$ matches with its
associated CFT entropy $S_{CFT}$ if one choose $P$ as
\begin{\eq}
P\approx \sqrt{d-2} ~8 \pi/\de E,
\end{\eq}
which is a universal constant, independent of $A$, as in the
three-dimensional case. But, in contrast to the three-dimensional
case, $c_{\eff}$ does not become a universal constant anymore,
\begin{\eq}
\label{c:AdS-Sch}
 c_{\eff}=c\approx \f{3 l^2 \de E}{\pi (d-1)
\sqrt{d-2}} \left(\f{\Om_{d-2}}{4 G}\right)^{1/(d-2)}
S_{BH}^{(d-3)/(d-2)}.
\end{\eq}
When the above results for the higher dimensional case of $d \geq
4$ are extrapolated to that of $d=3$, with an appropriate change
of definition of mass $M-\f{1}{8G} \ra M$,
the formula (\ref{c:AdS-Sch}), which becomes (\ref{clog}) exactly,
clearly shows that $c$ becomes a universal constant only for $d=3$
accidentally; this is in contrast to the speculation in Ref.
\ci{Carl:00}. (The reason will be explained in section 7.b.)

For the comparison with the $d=3$ case, the results for
$d=4~(\Om_2=4 \pi) $ are
\begin{\eq}
\label{adS-Sch4}
S_{CFT}&=&S_{BH}-\mbox{ln} S_{BH} +\f{1}{2}\mbox{ln}[ 4 \pi^2 l^4 (\de E)^2/(9 G)], \no \\
c_{\eff}&=&c\approx \f{ l^2 \de E}{\sqrt{2 \pi G}}  S_{BH}^{1/2},\no  \\
P&\approx& 8 \sqrt{2} \pi/\de E.
\end{\eq}

\section{Beyond the Logarithmic Corrections}

So far I have considered the logarithmic-order, including the
constant term as well, corrections, which are the first-order
corrections to the BH entropy. There I showed that the
holographic correspondence with the
 holographic screen at spatial infinity for the
BTZ black hole
has some disagreement.
However I showed that this mismatch can be resolved by
considering a horizon holography with an appropriate choice of the
period parameter $P$. Now, the important question would be of its
$consistency$: In the CFT manipulation at the horizon there is
only ``one'' undetermined parameter $P$ at the leading order, but
this was fixed by the requirement of the (horizon) holographic
principle even with the logarithmic-order corrections. Now then,
computing the higher order corrections beyond the logarithmic
order would be an important test of the consistency of the
holographic principle. So, the question is {\it whether  the $P$
determined at the logarithmic order has the perturbative
corrections, or that is enough to get a consistent (horizon)
holography, even with all the higher order corrections.}

To this end, let us first consider higher order terms in the
density of states $\Omega(E)$ of (\ref{Omega}) in the bulk. The
higher order terms in the sum beyond $n=2$, which is the
logarithmic-order that I have studied so far, start from $n=3$,
but this term seems to be problematic since its contribution
becomes $imaginary$ due to $(i \be')^3$ term. Actually, the
imaginary terms always appear for odd $n(\geq 3)$. But, due to the
reflection symmetry under $ \be'\lra -\be'$ of the relevant
integrand of (\ref{microZ}), as I have noted earlier (below
(\ref{Laplace})), these terms do not occur in our higher order
computations. So, the truly relevant terms start from $n=4$, and
these give the higher order corrections beyond the logarithmic
order ($n=2$). If I take into account these new corrections, one
can evaluate the integral of (\ref{Omega}) ``formally'' as follows
\begin{\eq}
\label{Omega2} \Om(E)= e^{\be E} Z[\be] \times \f{\de E}{\pi \sqrt
B_2} x^{1/2} e^x K_{1/4} (x) +(\mbox{h.o.t}),
\end{\eq}
such as entropy $S=\mbox{ln} \Om(E)$ becomes
\begin{\eq}
\label{S2} S&=&S_c +\mbox{ln} \left[ \f{\de E}{\pi \sqrt B_2}
x^{1/2} e^x K_{1/4}
(x) \right]+(\mbox{h.o.t}) \no \\
&=&S_c-\f{1}{2} \mbox{ln}\left[ \f{2 \pi B_2}{(\de E)^2}
\right]+\f{B_4}{8 (B_2)^2 }+O\left( \f{(B_4)^2}{(B_2)^4}\right),
\end{\eq}
where I have used the following formula in the first line
\begin{\eq}
\label{x^4}
 \int^{\infty}_{-\infty} d \be e^{-a \be^2 +b
\be^4}=\sqrt{\f{2}{a}} x^{1/2} e^x K_{1/4}(x)
\end{\eq}
with the modified Bessel function of the second kind $K_n(x)$ for
$x=-{a^2}/{(8 b)}$; in the second line, I have used the asymptotic
series expansion for a large value of $x=-{3 (B_2)^2}/{(4 B_4)}$
\begin{\eq}
\label{K}
 K_{1/4}(x)=\sqrt{\f{\pi}{2 x} }e^{-x} \left(1-\f{3}{32}
\f{1}{x} +O\left(\f{1}{x^2}\right)\right).
\end{\eq}
The (h.o.t)'s in (\ref{Omega2}) and (\ref{S2}) represent the
corrections from $\int^{\infty}_{-\infty} d \be ' \mbox{exp} [
\sum^{\infty}_{n\geq 5}  B_n (i \be')^n/ {n !}]$ and its
logarithm, respectively.
This asymptotic series expansion
might not be convergent for negative values of $x$, (i.e., $b<0$),
which seems to be a generic feature of black holes in AdS space as
in this paper, due to an essential singularity at $x=\infty$. But,
since \footnote{The combination $W=x^{1/2} e^x K_{n}(x)$ satisfies
the differential equation, $x^2 W'' +2 x^2 W'-(n^2-1/4)W=0$, while
$K_n(x)$ satisfies the modified Bessel's equation $x^2 K_n''+x
K_n'-(x^2+n^2)K_n=0$.}, the combination $x^{1/2} e^x K_{1/4}(x)$,
which appears in (\ref{x^4}), has no essential singularity at
$x=\infty$--actually there is no singularity at all--, the
infinite series of (\ref{K}) may be convergent even for the $b<0$
case, as well as the $b>0$ case, by Fuch's theorem; the
convergence of that combination for the $b<0$ case implies an
appropriate regularization prescription is required\footnote{ For
example,
$\lim_{x\ra \infty} \int^x_{-x} d \be$ instead of the integral of
(\ref{x^4}) }to get the finite answer from the integral
(\ref{x^4}), which is divergent $naively$;
 otherwise, the canonical ensemble is unstable $again$  for
 black holes in AdS space if one considers the higher order
 corrections, even though it is stable for the first-order
 correction, but this does not seem to occur \ci{Hawk:83,Abbo:82}.
 One might also like to continue this perturbative computation
to arbitrary higher orders, but this does not seems to be
straightforward since we do not know the integral formula for the
integrand $e^{\sum^{N}_{n=0} a_n \be^n}$ for $N
>4$ in (\ref{Omega}), which generalizes the formula of (\ref{x^4}); (\ref{S2}) is
the best correction as much as we can at present.

On the other hand, in the CFT side, one needs to compute
$\rho(\De)$ of (\ref{rho-expand}) by taking into account the
higher order terms in the sum  which start from $n\geq 3$
similarly to $\Om(E)$ computation; here also, one can safely
neglect the higher order correction terms in the bracket of
(\ref{rho-expand}), which is exponentially suppressed as $\tau_*
\ra i 0_+$. This higher order corrections may be evaluated by the
steepest descent method, which would produce a similar form as
(\ref{S2}). But, surprisingly an exact, convergent expansion, due
to Rademacher \ci{Rade:38}, exits, and I would like to use this
elegant formula for our purpose. Then, the formula reads
\ci{Dijk:00,Birm:00}, up to the exponentially suppressed terms,
which have been neglected also in the explicit steepest
computation,
\begin{\eq}
\label{rho:CFT4} \rho(\De) =e^{2 \pi \sqrt{c_{\eff} \De_{\eff}/6}}
\times \left(\f{c_{\eff}}{96 \De_{\eff}^3}\right)^{1/4} I_1(2 \pi
\sqrt{c_{\eff} \De_{\eff}/6})  +\mbox{(h.o.t.)}~,
\end{\eq}
such as
\begin{\eq}
\label{SCFT4} S_{CFT}&=&S_0+\mbox{ln}\left[\left(\f{c_{\eff}}{96
\De_{\eff}^3}\right)^{1/4}
I_1(S_0)   \right] +\mbox{(h.o.t.)} \no \\
&=&S_0+\mbox{ln}\left(\f{c_{\eff}}{96
\De_{\eff}^3}\right)^{1/4}-\f{3}{8} S_0^{-1} +O( (S_0)^{-2}),
\end{\eq}
where $I_n(x)$ is the modified Bessel function of the first kind,
and I have used its asymptotic series expansion for large $x$:
\begin{\eq}
\label{I}
 I_1(x)=\f{1}{\sqrt{2 \pi x}} e^x\left[1-\f{3}{8
} x^{-1}+O(x^{-2})\right].
\end{\eq}
The (h.o.t)'s in (\ref{rho:CFT4}) and (\ref{SCFT4}) would
represent the corrections from $\int_C d\tau~ \mbox{exp}
[\sum^{\infty}_{n\geq 5}  \eta^{(n)} (\tau-\tau_* )^n /{n !}]$ and
its logarithm, respectively.
Here, note that, similarly to (\ref{x^4}) case, (\ref{SCFT4}) can
be expressed as `$\mbox{ln}(\sqrt{x} e^x I_1(x)/ \sqrt{2
\De_{\eff}})+\mbox{(h.o.t.)}$' such as the infinite series of
(\ref{SCFT4}) may be convergent, though (\ref{I}) might not be in
general.\footnote{But note that, in contrast to (\ref{K}) case,
$x<0$ (i.e., $S_0<0$) is not allowed in this case, in order that
the saddle point approximation of section 3 works here.}

 Now, if I consider the CFT at the horizon,
(\ref{SCFT4}) becomes, using $c_{\eff},\De_{\eff}$ of
(\ref{horizonCFT}),
\begin{\eq}
\label{SCFT2'}
 S_{CFT}=S_{BH}-\f{1}{2} \mbox{ln} S_{BH}
-\mbox{ln} (\kappa P)-\f{3}{8} S_{BH}^{-1}+\f{1}{2} \mbox{ln} (8
\pi^3)+ O(S_{BH}^{-2}).
\end{\eq}
Now then, I am ready to test the consistency of our horizon
holography  by comparing the additional correction terms in
(\ref{S2}) and (\ref{SCFT2'}) with $P$ determined at the
logarithmic order. But, it is not difficult to show that there is
numerical mismatches with that $P$, such as the period $P$ also
get the perturbative corrections in order that the holographic
principle holds even for higher orders. To see this, I first note
that
\begin{\eq}
\f{B_4}{(B_2)^2}&=&-24( 15 J^8 l^8 +170 J^6 l^6 r_+^4 +1000 J^4
l^4 r_+^8 +480 J^2 l^2 r_+^{12}+128 r_+^{16})\no \\
&&\times [(J^2 l^2 -4 r_+^4)(
3 J^2 l^2 +4 r_+^4)^3 ]^{-1} \times S_{BH}^{-1} \no \\
&\approx &12 S_{BH}^{-1}~~~~~~~~~~~~~~~~~(\mbox{BTZ}), \\
\f{B_4}{(B_2)^2}&=&-[((-3+d)l^2+(-1+d)r_+^2)^5 (3-d+(-1+d)r_+^2
l^{-2})^2 ((-5+d)(-4+d)(-3+d)^4 l^8\no \\
&&+2(-3+d)^3(-1+d)(-40+11d)l^6 r_+^2-2(-3+d)^2(-1+d)^2
(-50+(-4+d)d)l^4r_+^4\no \\
&&-2(-3+d)(-1+d)^3(-4+11d)l^2 r_+^6 +(-1+d)^4d (1+d) r_+^8)) \no
\\
&&\times[(-2+d)l^8((-3+d)l^2-(-1+d)r_+^2)^5(-3+d+(-1+d)r_+^2
l^{-2})^6]^{-1} \times
 S_{BH}^{-1} \no \\
&\approx& \f{d(d+1)}{d-2} S_{BH}^{-1}
~~~~~~~~~~~~~~~~~(\mbox{Schwarzschild-AdS})~,
\end{\eq}
such as the bulk entropy (\ref{S2}) becomes
\begin{\eq}
\label{Sbulk2:BTZ} S&=&S_{BH}-\f{3}{2} \mbox{ln} S_{BH}
+\f{12}{8}S_{BH}^{-1}+\f{1}{2} \mbox{ln}[32 \pi^3 l^4 (\de E)^2
/(8 G)^2] +O(S_{BH}^{-2})~~~~~~(\mbox{BTZ}), \\
\label{Sbulk2:AdS-Sch} S&=&S_{BH}-\f{d}{2(d-2)} \mbox{ln} S_{BH}
+\f{d(d+1)}{8(d-2)}S_{BH}^{-1}+\f{1}{2} \mbox{ln}\left[\f{8  \pi
l^4 (\de E)^2}{(d-2)(d-1)^2} \left( \f{\Om_{d-2}}{4
G}\right)^{\f{2}{d-2}}
 \right] \no \\
 &+&O(S_{BH}^{-2})~~~~~~~~~~~~~~~~~~~~(\mbox{Schwarzschild-AdS})
\end{\eq}
for the BTZ and Schwarzschild-AdS black holes ($d \geq 4$),
respectively. Now then, by comparing (\ref{Sbulk2:BTZ}) and
(\ref{Sbulk2:AdS-Sch}) with (\ref{SCFT2'}),
one finds easily that they match when we re-normalize the period
as
\begin{\eq}
 P_{ren}&=&\f{8 \pi}{\de E} \left[ 1-\f{15}{8} S_{BH}^{-1}
+O(S_{BH}^{-2})\right]~~~~~~(\mbox{BTZ}),\\
P_{ren}&=&\f{\sqrt{d-2}~ 8\pi}{\de E} \left[
1-\f{d^2+4d-6}{8(d-2)} S_{BH}^{-1}
+O(S_{BH}^{-2})\right]~~~~~(\mbox{Schwarzschild-AdS})
\end{\eq}
for the BTZ and Schwarzschild-AdS black holes ($d \geq 4$),
respectively.

\section{Comparison with Other Approaches}

\label{WZW}
\subsection{Wess-Zumino-Witten(WZW) Model Approaches at the
Horizon}

\subsubsection{Lorenzian approach:} For the BTZ black
hole, there is alternative derivations of black hole entropy from
CFT at the horizon, whose details depend on the signature of the
metric, i.e., Lorenzian or Euclidean. I first consider the
Lorenzian approach in this section. In this case the CFT comes
from an induced $SL(2, {\bf R})\times SL(2, {\bf R})$ WZW model at
the black hole horizon, in the context of Chern-Simons formulation
\ci{Achu:86} of three-dimensional (Lorenzian) gravity, and the BH
entropy comes from the direct counting of its number of states in
the large $k=l/4G$ limit \ci{Carl:95}. There is an important
$qualilative$ difference between this and Ref. \ci{Carl:99} or
Ref. \ci{Stro:98}, which was analyzed in the previous sections:
The BH entropy $S_{BH}$ is a purely ``quantum'' effect, in
contrast to Ref. \ci{Carl:99} or Ref. \ci{Stro:98}, where the
$classical$ effect was dominant; but, one can not distinguish
between them at the leading order. However, quite interestingly,
it is known \ci{Carl:00} that {\it this WZW approaches gives the
correct `$-\f{3}{2} \mbox{ln} S_{BH}$' term already}, as is
consistent with the (horizon) holographic principle, though this
approach has some undesirable features of the non-unitary Hilbert
space and infinite degeneracy of the vacuum \ci{Carl:95}. So, the
interesting question is then whether this gives the higher order
correction terms $correctly$ as well, such as this may be
considered as the correct CFT candidate for a new AdS/CFT in the
context of a horizon holography. But, unfortunately this is not
the case, and the WZW model does not provide a correct holography
beyond the logarithmic order.

To see this, I first note that the associated partition function
becomes
\begin{\eq}
\label{Z:WZW}
 \hat{Z}_e [\tau]=\sum_{N} \rho(N) \mbox{exp}\left[ 2
\pi i \tau \left(N-\f{k^2 r_+^2}{l^2} \right) \right] (-i
\tau)^{-1/2}
\end{\eq}
with the density of states
\begin{\eq}
\label{rho:WZW}
 \rho(N)=\sqrt{\f{2k^2-1}{4 k^2}} \sum_{n=0}^{N}
\rho_0(N) \rho_0(N-n),
\end{\eq}
where $\rho_0$ is the density of states for an $SL(2,{\bf R})$ WZW
model \ci{Carl:00}. [Here, I am considering the modular
$non$-invariant partition function $\hat{Z}_e[\tau]$, following
Carlip. But, I have introduced $(-i \tau)^{-1/2}$ factor in
(\ref{Z:WZW}) as well as $(2 k^2-1)/(4 k^2)$ in (\ref{rho:WZW}),
which have been neglected in \ci{Carl:00}, to take into account
the integral for zero modes: $\int d \bar{\om} e^{2 \pi i \tau
(\De^++\De^-)}=\sqrt{\f{2 k^2-1}{-i \tau 4 k^2}} e^{-2 \pi i\tau
k^2r_+^2/l^2}$; actually $(-i \tau)^{-1/2}$ is just what is needed
to incorporate the extra factor in the transformation (A.2) of
Ref. \ci{Carl:00} into the our standard form (\ref{rho}).]

In the large $k$ limit, the three oscillators of $SL(2,{\bf R})$
can be treated independently, and (\ref{rho:CFT4}) gives
\begin{\eq}
\label{rho:sum}
 \rho(N) \approx \f{1}{8} \sum^N_{n=0}~e^{\sqrt{2} \pi (\sqrt{n} +\sqrt{N-n})}
 \ n^{-3/4} (N-n)^{-3/4}
 \left(1-\f{3}{8
\sqrt{2} \pi}\f{1}{\sqrt{n}} \right) \times \left(1-\f{3}{8
\sqrt{2} \pi}\f{1}{\sqrt{N-n}} \right)
\end{\eq}
up to $O(k^{-2})$ terms. Here, the last two terms came from the
second-order correction terms in (\ref{rho:CFT4}).

Furthermore, in the large $N$ limit, the sum in $\rho(N)$ may be
approximated as an integral
\begin{\eq}
\label{rho(WZW)}
 \rho(N)\approx \f{1}{8} \int^{\infty}_0 ~dx ~e^{\eta(x)}
 x^{-3/4} (N-x)^{-3/4}  \left(1-\f{3}{8 \sqrt{2} \pi}\f{1}{\sqrt{x}}
\right) \times \left(1-\f{3}{8 \sqrt{2} \pi}\f{1}{\sqrt{N-x}}
\right),
\end{\eq}
where
\begin{\eq}
\eta(x)=\sqrt{2} \pi (\sqrt{x} +\sqrt{N-x}).
\end{\eq}
Then, from the steepest descent method, similarly to the previous
sections, (\ref{rho(WZW)}) becomes
\begin{\eq}
 \rho(N) &\approx&  \f{2^{3/4}}{8} x_*^{-3/4} (N-x_*)^{-3/4} e^{\eta(x_*)}\times
\left(1-\f{3}{8 \sqrt{2} \pi}\f{1}{\sqrt{x_*}} \right) \times
\left(1-\f{3}{8 \sqrt{2} \pi}\f{1}{\sqrt{N-x_*}} \right) \no \\
  &\times& \int^{\infty}_0 ~dx ~\mbox{exp}
 \left\{ \f{1}{2} \eta^{(2)} (x-x_*)^2 +\sum^{\infty}_{n\geq
3} \f{1}{n !} \eta^{(n)} (x-x_* )^n \right\}\no \\
&=&  \f{1}{\sqrt{2 \sqrt{2} }} N^{-3/4}e^{ 2 \pi \sqrt{N} } \times
\left[ 1-\f{15+12 \sqrt{2}}{16 \pi} N^{-1/2} +O(N^{-1}) \right],
\end{\eq}
where
\begin{\eq}
\eta(x_*)=2 \pi \sqrt{N}
\end{\eq}
with $x_*=N/2$; here, interestingly, $\eta^{(4)} <0$, in contrast
to all other previous examples. Then, the CFT entropy at the
horizon becomes
\begin{\eq}
S_{}=\mbox{ln}~ \rho \approx 2\pi \sqrt{N} -\f{3}{2} \mbox{ln}
\sqrt{N}-\f{15+12 \sqrt{2}}{16 \pi} N^{-1/2}+O(N^{-1}).
\end{\eq}
Now, since $\hat{Z}_e[\tau]$ of (\ref{Z:WZW}) is dominated by the
state with
\begin{\eq}
\label{N} N=\f{k^2 r_+^2}{l^2} =\f{S_{BH}^2}{4 \pi^2},
\end{\eq}
which has been known as the physical state condition also
\ci{Carl:95}, the dominant terms in the entropy are
\begin{\eq}
\label{SWZW2} S_{}\approx S_{BH}-\f{3}{2}\mbox{ln} S_{BH}-\f{15+12
\sqrt{2}}{8}S_{BH}^{-1} +\f{1}{2} \mbox{ln}( 8 \pi^3)
+O(S_{BH}^{-2}),
\end{\eq}
which disagrees with the bulk entropy (\ref{Sbulk2:AdS-Sch}) at
the $S_{BH}^{-1}$ order by the factor of ``$-(5+4 \sqrt{2})/4$'',
as well as at the constant term.

Note that although the corrections due to the difference between
the sum (\ref{rho:sum}) and the integral (\ref{rho(WZW)}), when
$N$ is away from the infinity (but, $k$ is kept infinity still),
might affect the above result also, but one can check that this is
not the case since the effects are only the order of
$O(S_{BH}^{-3/2})$ from the Euler-Maclaurin integral formula.
(Details are omitted here.)

Hence, one finds that the WZW model approach for the
three-dimensional Lorenzian gravity does not favor the (horizon)
holography beyond the logarithmic order.
This might imply the breakdown of the equivalence between
three-dimensional gravity and the corresponding Chern-Simons
formulation beyond the logarithmic order since the WZW model
depends crucially on the Chern-Simons formulation; but, since
several other ingredients which are not directly related to the
formulation itself are also involved in this approach, this is not
clear at present.

\label{BTZ:Euc}
\subsubsection{Euclidean approach:}

As a complementary to the previous subsection, it would be also
interesting to compare it with the Euclidean continuation of the
black hole, which uses a quite different counting of states in
terms of the better-understood $SL(2,{\bf C})$ WZW model than the
poorly understood $SL(2, {\bf R})\times SL(2, {\bf R})$ WZW model
for the Lorenzian black holes
\ci{Carl:97}. It has been shown that these two different
approaches are actually related by a functional Fourier
transformation for the different boundary data at the horizons,
and they both give the BH entropy at the leading order
\ci{Carl:98}. The test of the equivalence beyond the leading order
is the subject of this subsection.

To this end, I note that, in the large $k=-l/4G$ limit, in which
$G$ is analytically continued to a negative value
\ci{Carl:98,Henn:91}, the canonical partition function for the
$SL(2,{\bf C})$ WZW model on the two-torus with modulus
$\tau=\tau_1+i \tau_2$
\begin{\eq}
\label{Z:sl}
 Z_{SL(2,{\bf C})}=|Z_{SU(2)}[\tilde{A}]|^2 \approx
\left|\mbox{exp} \left[ \f{\pi k}{4 \tau_2} \bar{u}^2 \right]
\bar{\chi}_{0k}(\bar{\tau},\bar{u})\right|^2
\end{\eq}
up to $O(k^{-1})$ terms, where $\tilde{A}_z$ on the horizon is
fixed to a constant value [$T_a=-i \sigma_2/2$;~ $\sigma_a$ are
Pauli's matrices]
\begin{\eq}
a=-\f{\pi i}{\tau_2} u T_3~,
\end{\eq}
and $\chi_{nk}$ are the Weyl-Kac characters for affine $SU(2)$
\ci{Elit:89}, which behaves asymptotically for large $\tau_2$
\begin{\eq}
\chi_{nk}(\tau,u)\approx \mbox{exp} \left\{ \f{\pi i}{2} \left[
\f{(n+1)^2}{k+2} -\f{1}{2} \right] \tau \right\}\f{\mbox{sin} \pi
(n+1) u}{\mbox{sin}\pi u}.
\end{\eq}
Then the usual number of states, following (\ref{rho}), for the
states in which the Virasoro generators $L_0$ and $\bar{L}_0$ have
eigenvalues $N$ and $\bar{N}$, respectively, becomes
\begin{\eq}
\rho(N,\bar{N})=-\f{1}{4 \pi^2} \int \f{d q_1}{q_1^{N-\bar{N}+1}}
\f{d q_2}{q_2^{N+\bar{N}-c/12+1}} Z_{SL(2,{\bf C})}(\tau),
\end{\eq}
where $q_1=e^{2 \pi i \tau_1},~q_2=e^{-2 \pi  \tau_2}$. Now, from
the physical state condition of $L_0-c/24=\bar{L}_0-c/24=0$
\footnote{In the original work \ci{Carl:97}, $Z_{SL(2,{\bf C})}$
was $implicitly$ assumed to be $Tr\{e^{2 \pi i \tau L_0} e^{-2 \pi
i \bar{\tau} \bar{L}_0}\}$, which is not modular invariant.
However, as can be easily checked, this is not correct since
(\ref{Z:sl}) is modular invariant \ci{Govi:01} as in our starting
partition function $Z[\tau, \bar{\tau}]$ of (\ref{CFTZ}). Hence,
the original work should be understood with a shift $L_0,\bar{L}_0
\ra L_0-c/24,\bar{L}_0-c/24$, respectively. Then, the final result
is consistent and the same as Carlip's eventually.} at the
horizon, the number of states at the horizon is given by
\begin{\eq}
\label{rho:sl}
 \rho\left(\f{c}{24}, \f{c}{24}\right)=i
\int^{\infty}_{-\infty} d \tau_1 \int^{\infty}_{-\infty} d \tau_2
~\mbox{exp} \left\{ \f{\pi k}{2} \left( \f{\Theta \tau_1}{2
\pi}-\f{|r_-|}{l} \right)^2- \f{\pi k}{2} \left[ \f{1}{\tau_2}
\left( \f{\Theta \tau_2}{2 \pi}+\f{r_+}{l} \right)^2-\f{2
\tau_2}{k+2} \right]\right\}.
\end{\eq}
This integral, after $\tau_1$ integration, can be evaluated by the
steepest descent method with the result
\begin{\eq}
 \rho\left(\f{c}{24},\f{c}{24}\right) \approx e^{\eta(\tau_{2*})} \sqrt{\f{2 \tau_{2*}}{k}}
 \f{2 \pi}{\Theta}  \int^{\infty}_{-\infty} ~d\tau_2 ~\mbox{exp}
 \left\{ \f{1}{2} \eta^{(2)} (\tau_2-\tau_{2*})^2 +\sum^{\infty}_{n\geq
3} \f{1}{n !} \eta^{(n)} (\tau_2-\tau_{2*} )^n \right\} ,
\end{\eq}
where
\begin{\eq}
\eta(\tau_2)=-\f{\pi k}{2} \left[\f{1}{\tau_2} \left(\f{\Theta
\tau_2}{2 \pi}+\f{|r_+|}{l} \right)^2 \right] +O(k^0),
\end{\eq}
which dominates the prefactor $\sqrt{{2 \tau_{2*}}/{k}}$, gets the
maximum
\begin{\eq}
\label{max:sl}
 \eta(\tau_{2*})&=&\f{2 \pi r_+}{4 G} \f{\Theta}{2
\pi} +O(k^0)
\end{\eq}
with
\begin{\eq}
\tau_{2*}&=&\f{r_+}{l} \f{\Theta}{2 \pi}+O(k^{-1})
\end{\eq}
when $r_+ \gg l$ is considered. Then, 
entropy $S=\mbox{ln} \rho(c/24,c/24)$ is computed as [$2
\pi-\Theta$ is the deficit angle of the conical singularity at the
horizon, and $\Theta$ is taken to be $2 \pi$ on-shell here.]
\begin{\eq}
S=S_{BH}+\f{1}{2} \mbox{ln} S_{BH} +\f{1}{2} \mbox{ln}\left(
\f{-4G}{\pi k l}\right)+\mbox{(h.o.t.)}.
\end{\eq}\
This show the disagreement, even at the logarithmic order, with
the bulk result (\ref{Sbulk:BTZ}), and also shows its
$in$equivalence with the Lorenzian approach of the previous
subsection beyond the leading order. Moreover, it is unclear, in
this result, whether or not the Bekenstein bound, i.e., $S<S_{BH}$
\ci{Beke:81} is satisfied since both $S_{BH}$ and $k$ are
extremely large.

Recently, it has been suggested to consider the modular invariant
density of states \ci{Govi:01} by employing the modular invariant
measure $d \tau_1 d \tau_2/(\tau_2)^2$ instead of the usual
measure $d \tau_1 d \tau_2$ in (\ref{rho:sl}), and in that
construction the correct logarithmic term `$-(3/2) \mbox{ln}
S_{BH}$', which agrees with the bulk result (\ref{Sbulk:BTZ}), was
obtained. But, its relevance to the usual thermodynamics is not
clear yet, which will be more discussed in the final section.
Moreover, even with that construction, the entropy disagrees with
the bulk result (\ref{Sbulk2:BTZ}) again beyond the logarithmic
order. (See Appendix {\bf A} for details.)

\subsection{Quantum Geometry Approach of the Horizon}

For four-dimensional non-rotating black holes with or without a
cosmological constant or electric, magnetic, and dilatonic
charges, there is an alternative derivation of black hole entropy
from the quantum geometry approach of Ashtekar et al. that counts
the boundary states of a three-dimensional Chern-Simons theory at
the horizon \ci{Asht:98,Asht:00,Kaul:00}. There were some
confusions about the logarithmic-correction term in this approach,
by identifying the entropy in this approach with the usual
(micro-canonical) entropy $S=\mbox{ln} \Om(E)$ and generalizing
this result in $d=4$ to other dimensions \ci{Carl:00}; as a result
of this, a wrong behavior of central charge $c$ (or $P$) in the
Cardy's formula (\ref{SCFT:ln}) has been speculated in Ref.
\ci{Carl:00}, as remarked in section 4. Clarification of this
issue is considered in this subsection, with the help of a recent
observation by Chatterjee and Majumdar \ci{Chat:03}.

To this end, I start by noting that the partition function in
quantum geometry approach is represented by
\begin{\eq}
\label{Z:qg}
 Z[\hat{\be}]=\sum_{n=0}^p {\cal N} (E(A_n))
e^{-\hat{\be} E(A_n)},
\end{\eq}
where $A_n$ is the area eigenvalue of the horizon with $n$
punctures each of which carries a spin $1/2$, and ${\cal
N}(E(A_n))$ is the degeneracy of the energy level $E(A_n)$
\ci{Asht:98,Kaul:00}. In the very large $p$ limit, this
may be approximated as an integral
\begin{\eq}
\label{Z':qg}
 Z[\hat{\be}]=\int_0^{\infty} dx~{\cal N} (E(A(x)))
e^{-\hat{\be} E(A(x))}.
\end{\eq}
Changing integration variable from $x$ to $E$ yields
\begin{\eq}
\label{Z':qg}
 Z[\hat{\be}]=\int_0^{\infty} dE \left|\f{dE}{dx} \right|^{-1}~{\cal N}(E)
e^{-\hat{\be} E}
\end{\eq}
such as, by comparing with the usual form of (\ref{Laplace}), one
obtains
\begin{\eq}
\Om (E)={\cal N} (E) \left|\f{dE}{dx} \right|^{-1} \de E.
\end{\eq}
Then, the usual micro-canonical entropy $S=\mbox{ln}\Om(E)$
becomes \ci{Chat:03}
\begin{\eq}
\label{S:qg}
 S=S_{QG}-\mbox{ln} \f{ |{dE}/{dx} |}{\de E},
\end{\eq}
where
\begin{\eq}
S_{QG}=\mbox{ln} {\cal N}(E(p)),
\end{\eq}
which is the entropy defined in quantum geometry approach
\ci{Asht:98,Asht:00,Kaul:00}. As we shall see, the additional term
in (\ref{S:qg}) is the key ingredient which resolves the above
mentioned confusions.

Now, from counting the number of conformal blocks of a
two-dimensional $SU(2)_{\hat{k}}$ WZW model with the level
$\hat{k}=A/(8 \pi \gamma G)$ that lives on the punctured 2-sphere
at $\hat{k} \ra \infty$ limit, one finds
\begin{\eq}
{\cal N}(E(p_0)) \approx \left( \begin{array} {c} p_0\\ p_0/2
 \end{array}\right)
-\left( \begin{array} {c} p_0\\ (p_0/2 -1)
 \end{array}\right)
\end{\eq}
for the largest number of punctures $p_0$ with all spins
$j_n=1/2$, which is given by
\begin{\eq}
\label{p0}
 p_0=\f{A}{4 G} \f{\gamma_0}{\gamma},
\end{\eq}
where $\gamma_0=1/(\pi \sqrt{3})$, and $\ga$ is the
Barbero-Immirzi parameter \ci{Barb:96}. Then, for large $p_0$, one
finds, from the asymptotic expansion formula of the factorial
function \footnote{ Note that the $p_0^{-1}$ and the constant
terms differ from Ref. \ci{Kaul:00}.},
\begin{\eq}
 S_{QG} &\approx& p_0 \mbox{ln}2-\f{3}{2} \mlo
p_0-\f{9}{4}
p_0^{-1} +\f{1}{2}\mlo ({8/\pi})  +O(p_0^{-2})  \no \\
&=&S_{BH} -\f{3}{2} \mlo S_{BH}-\f{9}{4} \mlo{2}
S_{BH}^{-1}+\f{1}{2}\mlo [{(\mlo 2)^3 8}/{ \pi}] +O(S_{BH}^{-2})
\label{S:QG}
\end{\eq}
by choosing $\ga=\ga_0 \mlo 2$ \ci{Asht:98,Kaul:00}\footnote{In
Ref.\ci{Carl:00}, $S_{QG}$ has been identified implicitly in
arbitrary dimensions as the micro-canonical entropy S. But, as we
have seen in (\ref{S:qg}), this is not correct because of the
second term, which can not be neglected. Furthermore, in higher
dimensions, the applicability of the formulation has not been
proved yet.}. Note that this result applies to the
four-dimensional Schwarzschild-AdS as well as the Schwarzschild
black holes since the formulation depends crucially on the local
properties of the horizon \ci{Asht:00}. However, when we compute
the usual micro-canonical entropy $S=\mlo \Om(E)$, there is a
sharp difference due to the additional term of (\ref{S:qg}).

To see this, let us first consider the four-dimensional
Schwarzschild-AdS black hole, in which the black hole mass $M$ is
given by, from
(\ref{M:adS-Sch}), 
\begin{\eq}
\label{M:adS-Sch4}
 2 G M =r_+ \left( 1 +\f{r_+^2}{l^2}
\right).
\end{\eq}
For a very large black hole in the presence of a cosmological
constant $\Lambda=-3/l^2$, i.e., $r_+ \gg l$,
this
reduces to
\begin{\eq}
M &\approx& \f{1}{2 G l^2} r_+^3 \no \\
&=&\f{1}{2 G l^2 (4 \pi)^{3/2}} A^{3/2},
\end{\eq}
where I have used $A=4 \pi r_+^2$ for the horizon area. With this
mass-area relation one finds
\begin{\eq}
\f{dE}{dx}=\f{dE}{dA} \f{dA}{dx}=\f{3}{2 (4 \pi)^{3/2} \sqrt{G}
l^2 } S_{BH}^{1/2} \f{dA}{dx}.
\end{\eq}
Now, by plugging $dA/dx \approx dA/dp_0=4 G \ga/\ga_0$, which is a
constant factor independent of the area $A$, one obtains finally
the microscopic entropy $S=\mlo \Om(E)$, from (\ref{S:qg}) and
(\ref{S:QG}),
\begin{\eq}
\label{S:adS-Sch:qg}
 S\approx S_{BH}
+\left(-\f{3}{2}-\f{1}{2}\right) \mlo S_{BH}-\f{9}{4} \mlo{2}~
S_{BH}^{-1}+\f{1}{2}\mlo \f{128 \pi^2 (\mlo 2) l^4 (\de E)^2}{9 G}
+O(S_{BH}^{-2}).
\end{\eq}
This disagrees with the bulk result (\ref{adS-Sch4}), even at the
logarithmic order. And, as was noted in Ref.\ci{Kaul:00} and
explicitly shown in Ref. \ci{Kaul:03}, the constant and the
$O(S_{BH}^{-1})$ terms might be affected by taking the level
$\hat{k}$ away from the asymptotic value ($\infty$), which we have
assigned above in the integral (\ref{Z:qg}), or by including the
spin values higher than $1/2$ such as we are away from the largest
number of punctures (\ref{p0}), but the logarithmic term is not
affected.

On the other hand, in the absence of a cosmological constant,
i.e., Schwarzschild black hole, (\ref{M:adS-Sch4}) reduces, for
$any$ black hole size, to
\begin{\eq}
M=\f{r_+}{2 G}=\f{1}{4 \sqrt{\pi} G} A^{1/2}
\end{\eq}
such as one finds, instead,
\begin{\eq}
\label{Sch4}
 S\approx S_{BH} +\left(-\f{3}{2}+\f{1}{2}\right) \mlo
S_{BH}-\f{9}{4} \mlo{2}~ S_{BH}^{-1}+\f{1}{2}\mlo [128 \mlo 2 ~G
(\de E)^2] +O(S_{BH}^{-2}).
\end{\eq}
But, unfortunately, there is no comparable bulk entropy
computation in this case since the canonical partition function
(\ref{Laplace}) diverges such as the expansion of $\Om(E)$ as in
(\ref{Omega}) is not defined, though, interestingly, this has the
same logarithmic term of (\ref{Sbulk:AdS-Sch}) for $d=4$: We need
an independent framework to compute $\Om(E)$; a possible way would
be to consider the canonical ensemble of the black hole within a
finite cavity \ci{York:86}, though it is too artificial
\ci{Hawk:83}, but because of the finite size of the cavity
compared to the Schwarzschild radius $r_+=2 M$, the result can not
be expanded as in (\ref{Sch4}) \ci{Akba:03}; we need an estimation
in the micro-canonical ensemble from first principles. In this
circumstance, it would be interesting to compare with the
horizon-CFT approach, which applies also to the Schwarzschild
black hole. In this approach, one obtains the microscopic entropy,
from (\ref{SCFT:ln}) and (\ref{SCFT4}), as
\begin{\eq}
\label{SCFT-Sch}
 S_{CFT}=S_{BH}-\mbox{ln} P
-\f{3}{8} S_{BH}^{-1}+\f{1}{2} \mbox{ln} (32 G)+ O(S_{BH}^{-2}),
\end{\eq}
where I have used
\begin{\eq}
\kappa =\f{\pi}{4 G M}=\f{\pi^{3/2}}{2 \sqrt{G}} S_{BH}^{-1/2}.
\end{\eq}
Note that ``$-(1/2) \mbox{ln} S_{BH}$'' term cancels `$-\mbox{ln}
\kappa\sim (1/2) \mbox{ln} S_{BH}$' in (\ref{SCFT:ln}), such as
only ``$-\mbox{ln} P$'' term remains in (\ref{SCFT-Sch}). The
quantum geometry result (\ref{Sch4}) would agree with this result
if I choose
\begin{\eq}
P=a S_{BH} +b
\end{\eq}
with the appropriate constants $a$ and $b$; $a^2=1/(4 \mbox{ln} 2
(\de E)^2) , b=((9/4) \mbox{ln} 2-(3/8)) a$. But, this would be
quite questionable since there is the disagreement in the
Schwarzschild-AdS case always, as already was noted in
(\ref{S:adS-Sch:qg}), and there is no resolution in the present
context yet; a modification in the classical action of
Ref.\ci{Asht:98} might be needed for the resolution, but it not
clear at present.

In summary for quantum geometry approach, there is the
disagreement beyond the logarithmic term with the bulk result of
Schwarzschild-AdS black holes, but there is no way to test the
holography for the Schwarzschild black hole case, i.e., Flat/CFT
due to the absence of its corresponding bulk entropy computation.

\section{Conclusion and Open Questions}

I have tested the holographic principle for mainly AdS spaces by
examining the logarithmic and higher order corrections to the
Bekenstein-Hawking entropy of black holes. For the BTZ black hole,
the
holographic correspondence with the screen
 at spatial infinity
has some disagreement
 beyond the logarithmic correction ( by the
factor of $2$ at the logarithmic order), which may be
in contrast to
the
AdS/CFT
predictions.
On the other hand, a holography at the event horizon
from Carlip's horizon-CFT approach of black hole entropy in any
dimension \ci{Carl:99,Park:99,Park:02} is more flexible, and can
be made to be satisfied by choosing an appropriate  period
parameter $P$, which has been arbitrary for the leading-order
computation. The analysis on the higher dimensional Schwarzschild
black holes in AdS space shows a universality of the choice $P$,
except for a dimension-dependent factor. I have also compared with
several other approaches, i.e., the induced WZW model approaches
at the horizon for the Lorenzian and the Euclidean BTZ black
holes, and also the quantum geometry approach of Ashtekar et al.
at the horizon for the four-dimensional Schwarzschild and
Schwarzschild-AdS black holes. In these analysis I have shown that
none of these other approaches satisfies the (horizon) holographic
principle, in contrast to our horizon-CFT approach. As a
byproduct, I have clarified some confusions in the quantum
geometry approach which result from some misunderstandings of the
definition of entropy. However, there remain several open
questions, which are listed below.

1. {\it Quantum corrections} \footnote{I thank for S. Deser, D. V.
Fursaev, J. Maldacena, and S. N. Solodukhin for discussion.}:
While I have used the expansion (\ref{Omega}) around $\beta'=0$, I
have not touched $Z[\beta]$, which
may include
 all quantum loop corrections by
$Z[\beta]=Tr(exp(-\beta H))=\int d[g] exp(iI[g])$
\ci{Gibb:77}. If we consider the loop corrections of the
gravitational fields we need to generalize the Einstein gravity to
a higher curvature gravity \ci{tHoo:74}--in this case, the
canonical entropy (\ref{Sc}) is not simply the BH form
(\ref{SBH}), but also includes a sum of curvature invariants
integrated over the horizon \ci{Jaco:93} \footnote{The higher
curvature effect in the black hole entropy would be also
reproduced by the Cardy's formula with the corresponding
corrections in the central charge and $L_0$ eigenvalue. See
\ci{Civi:03} for this possibility.}-- with a $possible$ UV-finite
$ln A$ correction. But this would not affect $much$ our result for
large black holes from the following reasons:

i). In this paper, I have considered mainly the black holes in AdS
space, which is stable under the fluctuations of thermal gravitons
\ci{Hawk:83}. In this case, it seems that the usual $ln A$
correction from some ambiguity in the renormalization for the d=4
Schwarzschild black hole, due to lack of a natural length scale
$\mu$ other than $r_{+}=2M l^2_{Planck}$
\ci{Furs:94,Furs:95,Solo:98}, would not appear \ci{Wins:01} due to
another natural length scale $l \sim \sqrt{-\Lambda}$ in AdS. All
the other terms are UV-divergent which can be renormalized away.

ii). Now, involving the higher curvature terms in entropy, which
is renormalized, these would be very small for large black holes,
as in this paper, since the curvature near the horizon decreases
as the horizon area $A$ increases. (For more details, see the
argument by Carlip in Appendix B of \ci{Carl:00}.) Moreover, for
three dimensions, in particular, one can expect no gravitational
loop correction due to the absence of propagating graviton
\ci{Witt:88}, such as our results for the BTZ black hole in
section 4 are not affected by quantum corrections, except for some
possible finite terms which would be renormalized away
\ci{Carl:97}. However, an explicit computation of those higher
curvature terms in the entropy for the CFT as well as the bulk
would be quite interesting.

2. {\it De-Sitter space}: From the empirical reason, holography
for de-Sitter space would be more interesting, but the situation
is not so affirmative as that of AdS space: The only thing I can
show (the details will be appeared elsewhere \ci{Park:04}) is
that, by generalizing the steepest descent method for
``imaginary'' $\De$ and $c$ , the KdS$_3$ solution \ci{Park:98}
has the same factor of 2 discrepancy at the logarithmic order for
the entropy at spatial infinity as the BTZ black hole; this shows
clearly  the usual dS$_3$/CFT$_2$ $might$ not be correct either
beyond the logarithmic correction. But, when we consider the
entropy at the cosmological horizon, which is the only horizon in
this case, the steepest descent approximation is questionable
since the convergence of an expansion around the steepest descent
path is not guaranteed, which is essentially due to finite
$r_+/l$; this problem is similar to the situation for the black
holes in a finite cavity \ci{York:86}. We need other ways to
compute the entropy corrections such that one can get a convergent
expansion around the steepest descent path  with some controlling
parameters.

3. {\it Flat space}: Can we ``directly'' compute the bulk density
of states $\Om(E)$ for the asymptotically flat Schwarzschild black
holes, without recourse to the Laplace transformation of the
canonical partition function (\ref{microZ}), which is divergent in
this case ? This can be used for testing F(lat)/CFT
correspondence, which has been recently considered
\ci{Bous:99,Son:02}.

4. {\it Modular invariance}: In the CFT side of section 3, the
modular invariance--especially the invariance under $\tau \ra
-1/\tau$--of the partition function $Z[\tau, \bar{\tau}]$ in
(\ref{CFTZ}) has been a crucial role in determining its density of
states $\rho(\De, \bar{\De})$ of (\ref{rho-ser}) perturbatively.
But, this property is clearly absent in the usual canonical
partition function in section 2, which implies some mismatches
between $\Om(E)$ in (\ref{microZ}) and $\rho(\De,\bar{\De})$ in
(\ref{rho}). One might consider the modular invariant measure as
in \ci{Govi:01, Kuta:91} in order $not$ to double-count the states
which are connected by the modular transformations. But, in this
case the usual connection of $\tau \sim i \be/l$ \ci{Mald:98}
would be lost
 for large $\be$ (i.e., small temperature $T$) case.
A possible resolution might be to generalize the usual
thermodynamics relations in section 2 such that a similar duality
in the thermal temperature $T$ is favored as in Ref.\ci{Dien:03}.
Even with this unclear circumstance in the large $\be$, our main
computation for small $\be$ (i.e., high $T$) black holes seems to
be quite safe still. But, for an affirmative answer, we need to
compute directly the density of states $\rho(\De, \bar{\De})$
without recourse to the modular invariance. Is there any way to do
this ?

5. {\it String theory}: In string theory side, similar
disagreements in two-loop tests of the PP-wave/YM correspondence
have been reported \ci{Beis:03,NKim:03}, and there has been no
clear resolution yet. The disagreement are the factor of $2$ for
Ref.\ci{Beis:03}, and approximately 2 (i.e., 19/8) for Ref.
\ci{NKim:03}; for the latter case, a resolution was proposed in
the later paper \ci{NKim:03}, but its independent confirmation
seems to be unclear so far. This factor of 2 disagreement
resembles the same factor disagreement for the BTZ black hole
entropy at the logarithmic order.
Is this just an accidental coincidence, or is there any deeper
reason ?

6. {\it Much higher order corrections}: In the evaluation of much
higher order corrections $(n > 4)$ for $\rho(\De, \bar{\De})$ we
can use the exact formula of Rademacher to any order, which
implies the integral with an infinite sum in the exponent of the
integrand in (\ref{rho-expand}) can be computed. Then, can we
obtain, from the Rademacher's exact formula, the integral formula
for $\int^{\infty}_{-\infty} dx~{\mbox exp} \{\sum_{n=2}^{N} a_n
x^n \}$ for arbitrary $N > 4$ such that the integral in $\Om(E)$
of (\ref{Omega}) can be computed to arbitrary orders also ? This
would be an interesting problem in mathematics itself also.

7. {\it Chern-Simons theory versus gravity}: It is well known that
the three-dimensional Chern-Simons action with an appropriate
gauge group is equivalent to the three-dimensional pure gravity
action ``on-shell'' \ci{Achu:86}, and this is true even for the
usual boundary term  ``$\oint 2 K$'' at spatial infinity as well
as the bulk Einstein-Hilbert action \ci{Bana:98}. Then, does our
result in section 7.a imply that this equivalence is not true at
quantum level, which is ``off-shell'' necessarily, or the boundary
term of the Chern-Simons action for the horizon should be modified
in accordance with the appropriate horizon term in gravity side
\ci{Park:02} ?

Note added: While this paper was being written, a paper
\ci{Myun:04} appeared which noted also the factor of 2 mismatch
for the BTZ black hole between the bulk and Strominger's
logarithmic correction to the BH entropy. But there, in contrast
to my paper, this mismatch was neglected and not seriously
considered .

\begin{center} {\bf Acknowledgments}
\end{center}
First of all, I appreciate Prof. Yon-Soo Myung for drawing my
attention to Ref. \cite{Das:02}, helpful discussions at the early
stage, and for encouraging me during this work. I appreciate
Profs. Niklas Beisert, Steve Carlip, Stanley Deser, Dmitri V.
Fursaev, Kerson Huang, Nakwoo Kim, Parthasarathi Majumdar, Juan
Maldacena, Robert Mann, John Schwarz, Kostas Skenderis, Segrey N.
Solodukhin, Suneeta Vardarajan, Anastasia Volovich for helpful
correspondences. I also appreciate Prof. Jaemo Park for
hospitality while I have been staying at POSTECH. This work was
supported by the Korean Research Foundation Grant
(KRF-2002-070-C00022).

\appendix

\section{}

In this appendix, I compute the black hole entropy of Euclidean
black holes of section 7 a.2 
for the modular invariant density of states, as suggested by
Govindarjan et al. \ci{Govi:01}. The suggested form of the density
of states corresponds to a modified density of states, in contrast
to the usual $\rho(c/24, c/24)$ of (\ref{rho:sl}),
\begin{\eq}
\label{rho':sl}
 \tilde{\rho}\left(\f{c}{24}, \f{c}{24}\right)=i
\int^{\infty}_{-\infty} d \tau_1 \int^{\infty}_{-\infty} d \tau_2
\f{1}{(\tau_2)^2} Z_{SL(2,{\bf C})}(\tau)
\end{\eq}
with the modular invariant measure, by incorporating
$1/(\tau_2)^2$ factor, which is well-known factor in string theory
context \ci{Kuta:91}. Here, since $Z_{SL(2,{\bf C})}(\tau)$ is
modular invariant by itself on a two-torus, the density of states
$\tilde{\rho}$ itself is modular invariant also; of course, the
density of states for the higher modes $\rho(N,\bar{N})$ is not
modular invariant still, even with the $1/(\tau_2)^2$ factor, and
needs a more generalized factor in that case.

The integral (\ref{rho':sl}) can be evaluated by the steepest
descent method as in (\ref{rho-expand})
\begin{\eq}
\label{rho'-ser}
 \tilde{\rho}\left(\f{c}{24},\f{c}{24}\right) &\approx&  \f{e^{\eta(\tau_{2*})}}{(\tau_{2*})^2} \sqrt{\f{2 \tau_{2*}}{k}}
 \f{2 \pi}{\Theta}  \int^{\infty}_{-\infty} ~d\tau_2 ~\mbox{exp}
 \left\{ \f{1}{2} \eta^{(2)} (\tau_2-\tau_{2*})^2 +\sum^{\infty}_{n\geq
3} \f{1}{n !} \eta^{(n)} (\tau_2-\tau_{2*} )^n \right\} \no \\
&&\times \left[ 1+\sum^{\infty}_{m \geq 1} \f{1}{m !}(\tau_{2*})^2
[(\tau_{2*})^{-2}]^{(m)}|_{\tau_{2*}} (\tau_2-\tau_{2*})^m
\right],
\end{\eq}
where $\eta(\tau_2)=-\f{\pi k}{2} [\f{1}{\tau_2} (\f{\Theta
\tau_2}{2 \pi}+\f{|r_+|}{l} )^2 ] +O(k^0)$, which dominates the
prefactor $(\tau_{2*})^{-2} \sqrt{{2 \tau_{2*}}/{k}}$, gets the
same maximum as (\ref{max:sl}) when $r_+ \gg l$ is considered;
there are no terms of odd power in $(\tau_2-\tau_{2*})$ in the
exponent of (\ref{rho'-ser}), due to the same reason for
(\ref{rho-expand}) case. Then, from the integral formula
(\ref{x^4}) and the asymptotic form of the modified Bessel
function of the second kind, one obtains black hole entropy
$\tilde{S}\equiv \mbox{ln} \tilde{\rho}(\f{c}{24},\f{c}{24})$ as
follows:
\begin{\eq}
\tilde{S}=S_{BH}-\f{3}{2} \mbox{ln} S_{BH} +\f{24}{8}
S_{BH}^{-1}+\f{1}{2} \mbox{ln}\left( \f{\pi^3
l^2}{G^2}\right)+O(S_{BH}^{-2},k^{-1}).
\end{\eq}
Note that one obtains the correct `$-3/2$' factor in the
logarithmic term, which agrees with the bulk result
(\ref{Sbulk:BTZ}) \ci{Govi:01}: This is essentially due to the
additional factor $1/(\tau_2)^2$ in the measure, which ``reduces''
the counted number of states in general [`$-2 \mbox{ln} S_{BH}$'
in our case] by removing the over-counted states due to the
modular transformation. However, even in this construction, the
higher order terms, which are actually the same as in the case
without the $1/(\tau_2)^2$ factor, disagrees with the bulk result
(\ref{Sbulk2:BTZ}) again.


\begin{thebibliography}{999}

\bibitem{Hoof:93} G. 't Hooft, {\it in} Salem Fest (World
Sceintific Co., Singapore, 1993); L. Susskind, J. Math. Phys. {36}
(1995) 6377.

\bb{Ahar:99} O. Aharony, S. S. Gubser, J. Maldacena, H. Ooguri, Y.
Oz, Phys. Rept, 323 (2000) 183
and references therein.

\bibitem{Park:98} M.-I. Park, Phys. Lett. {B 440} (1998);
275; 
Nucl. Phys. {B 544} (1999) 377.
\bb{Bous:99}
 R. Bousso, JHEP {9906} (1999) 028. 
\bb{Stro:01}
 A. Strominger, JHEP {0110} (2001) 034; 
 M. Spradlin, A. Strominger, and A. Volovich,
hep-th/0110007.

\bb{Son:02} D. T. Son and A. O. Stainets, JHEP 0209 (2002) 042;
J. de
Boer and S. N. Solodukhin, Nucl. Phys. B 665 (2003) 545; 
 G. Arcioni and C.
Dappiaggi, Nucl. Phys. B 674 (2003) 553;
hep-th/0312186.

\bb{Sach:01} I. Sachs and S. N. Solodukhin, Phys. Rev. {D 64}
(2001) 124023.

\bb{Carl:95} S. Carlip, Phys. Rev. {D 51} (1995) 632.

\bb{Carl:97} S. Carlip, Phys. Rev. {D 55} (1997) 878.

\bibitem{Asht:98} A. Ashtekar, J. Baez, A. Corichi, and K. Krasnov,
  Phys. Rev. Lett. {80} (1998) 904.

\bibitem{Teit:96} C. Teitelboim, Phys. Rev. {D 53} (1996) 2870.

\bibitem{Carl:99} S. Carlip, Phys. Rev. Lett. {82} (1999) 2828;
ibid. {83} (1999) 5596.

\bb{Park:99} M.-I. Park and J. Ho, Phys. Rev. Lett.{83} (1999)
5595; M.-I. Park and J. H. Yee, Phys. Rev. {D 61} (2000) 088501.

\bibitem{Park:02} M.-I. Park, Nucl. Phys. {B 634} (2002) 339.

\bibitem{Solo:99} S. N. Solodukhin, Phys. Lett. {B 454} (1999) 213.

\bibitem{Carl:02} S. Carlip, Class. Quant. Grav. {16} (1999) 3327;
 Phys. Lett. {B 508} (2001) 168; Phys. Rev. Lett. {88} (2002) 241301.

\bb{Kang:04} G. Kang, J.-i. Koga, and M.-I. Park, hep-th/0402113.

\bb{Stro:98} A. Strominger, JHEP, {02} (1998) 009.

\bibitem{Hawk:76} S. W. Hawking, Phys. Rev. {D 13} (1976) 191.

\bibitem{Brow:94} J. D. Brown and J. W. York, Phys. Rev. {D 47}
(1993) 1420; gr-qc/9405024.

\bb{Reif:86} F. Reif, {Fundamentals of Statistical and Thermal
Physics} (McGraw-Hill Book Co., 1986); M. Plischke and B.
Bergersen, {Equilibrium Statistical Physics} (Prentice-Hall Inc.,
New Jersey, 1989).

\bb{Huan:63} K. Huang, {Statistical Mechanics} (John Wiley \&
Sons., Singapore, 1963 ); G. Morandi, {Statistical Mechanics}
(World Scientific, Singapore, 1995).

\bibitem{Gibb:77} G. W. Gibbons and S. W. Hawking, Phys. Rev. {D 15}
(1977) 2752.

\bibitem{Brow:91} J. D. Brown, E. A. Martinez, and J. W. York,
  Phys. Rev. Lett. {66} (1991) 2281.

\bb{Das:02} S. Das, P. Majumdar, and R. K. Bhadur, Class. Quantum.
Grav. {19} (2002) 2355; Y. S. Myung, ibid, 21 (2004) 1279.

\bibitem{Card:86} J. A. Cardy, Nucl. Phys. {B 270} (1986) 186.

\bibitem{Carl:98} S. Carlip, Class. Quant. Grav. {15} (1998) 3609.

\bb{Park:04} M.-I. Park, in preparation.

\bibitem{Dese:84} S. Deser, R. Jackiw, and G. 't Hooft, Ann. Phys.
(N.Y.) {152} (1984) 220; A. R. Steif, Phys. Rev. {D 53} (1996)
5527.

\bibitem{Bana:92} M. Ba\~nados, Teitelboim and J. Zanelli, Phys. Rev.
Lett. {69} (1992) 1849.

\bibitem{Hawk:83} S. W. Hawking and D. Page, Commun. Math. Phys.
{87} (1983) 577; S. W. Hawking, C. J. Hunter, and M. M.
Taylor-Robinson, Phys. Rev. {D 59} (1999) 064005.

\bb{Furs:95} D. V. Furasev, Phys. Rev. {D 51} (1995) 5352.

\bb{Frol:96} V. P. Frolov, W. Israel, and S. N. Solodukhin, Phys.
Rev. {D 54} (1996) 2732.

\bb{Mann:98} R. B. Mann and S. N. Solodukhin, Nucl. Phys. {B 523}
(1998) 293.

\bb{Solo:98} S. N. Solodukhin, Phys. Rev. {D 57} (1998) 2410.

\bb{Suss:93} L. Susskind, hep-th/9309145.

\bb{Brow:86} J. D. Brown and M. Henneaux, Commun. Math. Phys. 104
(1986) 207.

\bb{Oh:98} P. Oh and M.-I. Park, Mod. Phys. Lett. A 14 (1998) 231.

\bb{Park:0403} M.-I. Park, hep-th/0403089.

\bb{Carl:00} S. Carlip, Class. Quant. Grav. 17 (2000) 4175.

\bb{Kaul:00} R. K. Kaul and P. Majumdar, Phys. Rev. Lett. {84}
(2000) 5255.

\bb{Henn:85} M. Henneaux and C. Teitelboim, Commin. Math. Phys.
{98} (1985) 391.

\bb{Hyun:97} S. Hyun, J. Korean Phys. Soc. 33 (1998) S532;
K. Sfetsos and K. Skenderis,
Nucl. Phys. {B 517} (1998) 179.

\bb{Bana:00} M. Banados, A. Chamblin, and G. W. Gibbons, Phys.
Rev. {D 61} (2000) 081901; Brecher, A. Chamblin, and H. S. Reall,
Nucl. Phys. B607 (2001) 155.

\bb{Jing:00} J. Jing and M.-L. Yan, Phys. Rev. {D 63} (2000)
024003.

\bibitem{Myer:86} R. C. Myers and M. J. Perry, Ann. Phys. {172}
(1986) 304.

\bb{Asht:84} A. Ashtekar and A. Magnon, Class. Quantum Grav. {1}
(1984) L39; A. Ashtekar and S. Das, {\it ibid.}, {17} (2000) L17.

\bb{Abbo:82} L. F. Abbott and S. Deser, Nucl. Phys. {B 195} (1982)
76; G. W. Gibbons, S. W. Hawking, G. W. Horowitz, and M. J. Perry,
Commun. Math. Phys. {88} (1983) 295.

\bb{Rade:38} H. Rademacher, Amer. J. Math. {60} (1938) 501.

\bb{Dijk:00} R. Dijkgraaf, J. Maldacena , G. Moore, and E.
Velinde, hep-th/0005003.

\bb{Birm:00} D. Birmingham and S. Sen, Phys. Rev. D 63 (2001)
047501.

\bb{Achu:86} A. Achu'caro and P. K. Townsend, Phys. Lett. {B 180}
(1986) 89; E. Witten, Nucl. Phys. {B 311} (1988) 46.

\bb{Henn:91} M. Henningson, S. Hwang, and P. Robert, Phys. Lett.
{B 267} (1991) 350.

\bb{Elit:89} S. Elitzur et al., Nucl. Phys. {B 326} (1989) 108; J.
M. F. Labastida and A. V. Ramallo, Phys. Lett. {B 227} (1989) 92;
E.Witten, Commun. Math. Phys. {137} (1991) 29; F. Falceto and K.
Gawedzki, Commun. Math. Phys. {159} (1994) 549 ; N. Hayashi, Prog.
Theor. Phys. Suppl. {114} (1993) 125.

\bb{Govi:01} T. R. Govindarajan, R. K. Kaul, and V. Suneeta,
Class. Quantum. Grav. {18} (2001) 2877.

\bb{Beke:81} J. D. Bekenstein, Phys. Rev. {D 23} (1981) 287; { 49}
(1994) 1912.

\bibitem{Asht:00} A. Ashtekar, A. Corichi, and K. Krasnov,
  Adv. Theor. Math. Phys. {3} (2000) 419. 

\bb{Chat:03} A. Chatterjee and P. Majumdar, gr-qc/0309026.

\bb{Barb:96} G. Barbero and J. Fernando, Phys. Rev. {D 54} (1996)
1492; G. Immirzi, Nucl. Phys. (Proc. Suppl.) {57} (1997) 65.

\bb{Kaul:03} R. K. Kaul and S. K. Rama, Phys. Rev. {D 68} (2003)
024001.

\bb{York:86} J. W. York, Phys. Rev. {D 33} (1986) 2092.

\bb{Akba:03} M. M. Akbar and S. Das, hep-th/0304076.

\bb{tHoo:74} 't Hooft and M. Veltman, Ann. Inst. H. Poincare {A
20} (1974) 69; M. Goroff and A. Sagnotti, Nucl. Phys. {B 266}
(1986) 799; A. E. M. van de Ven, Nucl. Phys. {B 378} (1992) 309.

\bb{Jaco:93} T. Jacobson and R. C. Myers, Phys. Rev. Lett. {70}
(1993) 3684; T. Jacobson, G, Kang, and R. C. Myers, Phys. Rev.
{D~49} (1994) 6587; ibid. {D52} (1995) 3518.

\bb{Civi:03} M. Civitan, S. Pallua, P. Prester, Phys. Lett. {B
546} (2002) 119; ibid. {B 555} (2003) 248.

\bb{Furs:94} D. V. Fursaev and S. N. Solodukhin, Phys. Lett. {B
365} (1996) 51.

\bb{Wins:01} E. Winstanley, Phy. Rev. {D 63} (2001) 084013.

\bb{Witt:88} E. Witten, Nucl. Phys. {B 311} (1988/89) 46; S.
Deser, J. McCarthy, and Z. Yang, Phys. Lett. {B 222} (1989) 61.

%

\bb{Kuta:91} D. Kutasov and N. Seiberg, Nucl. Phys. {B 358} (1991)
600.

\bb{Mald:98} J. Maldacena and A. Strominger, JHEP 9812 (1998) 005.

\bb{Dien:03} K. R. Dienes and M. Lennek, hep-th/0312173.

\bb{Beis:03} N. Beisert, C. Kristjansen, and M. Staudacher, Nucl.
Phys. B664 (2003) 131;
M. Spradlin and A. Volovich, hep-th/0303220.

\bb{NKim:03} N. Kim, T. Klose, and J. Plefka, Nucl. Phys. B 671
(2003) 359;
T. Klose and J. Plefka, Nucl. Phys. B 679 (2004) 127.

\bb{Bana:98} M. Ba$\tilde{n}$ados and F. Me'ndez, Phys. Rev. {D
58} (1998) 104014.

\bb{Myun:04} Y. S. Myung, Phys. Lett. B579 (2004) 205.
\end{thebibliography}
\end{document}